\pdfoutput=1

\documentclass[aip,jcp,reprint,groupedaddress,footinbib]{revtex4-1}
\usepackage{graphicx} %
\usepackage{amsmath}
\usepackage{amssymb} %
\usepackage{bm}
\usepackage{comment}
\usepackage{tabularx}
\usepackage{booktabs}

\usepackage{siunitx}
\DeclareSIUnit{\calorie}{cal}
\DeclareSIUnit{\Calorie}{\kilo\calorie}
\DeclareSIUnit{\atmass}{amu}
\usepackage{subfig}

\usepackage{hyperref}
\hypersetup{colorlinks=true,linkcolor=blue,citecolor=blue,urlcolor=blue}

\usepackage{color}

\usepackage{braket} %
\newcommand{\ketbra}[2]{\ket{#1}\!\bra{#2}}

\newcommand{\eu}{\mathrm{e}^}
\newcommand{\iu}{\ensuremath{\mathrm{i}}}
\newcommand{\rmd}{\mathrm{d}}
\newcommand{\half}{{\ensuremath{\frac{1}{2}}}}
\newcommand{\thalf}{{\ensuremath{\tfrac{1}{2}}}}

\newcommand{\op}[1]{\ensuremath{\hat{#1}}}

\renewcommand{\mathbf}[1]{\bm{#1}}

\DeclareMathOperator{\Tr}{Tr}

\newcommand{\der}[3][]{\frac{\rmd^{#1}{#2}}{\rmd{#3}^{#1}}}
\newcommand{\pder}[3][]{\frac{\partial^{#1}{#2}}{\partial{#3}^{#1}}}
\newcommand{\pders}[3]{\frac{\partial^2{#1}}{\partial{#2}\partial{#3}}}

\newcommand{\eqn}[1]{Eq.\,(\ref{#1})}

\newcommand{\eqs}[1]{Eqs.\,(\ref{#1})}

\newcommand{\Ref}[1]{Ref.~\onlinecite{#1}}
\newcommand{\Refs}[1]{Refs.~\onlinecite{#1}} %

\usepackage{dcolumn} %
\newcolumntype{d}[1]{D{.}{.}{#1}} %

\newcolumntype{L}[1]{>{\raggedright\arraybackslash}p{#1}} %
\newcolumntype{C}[1]{>{\centering\arraybackslash}p{#1}} %
\newcolumntype{R}[1]{>{\raggedleft\arraybackslash}p{#1}} %

\DeclareMathOperator*{\SumInt}{%
\mathchoice%
  {\ooalign{$\displaystyle\sum$\cr\hidewidth$\displaystyle\int$\hidewidth\cr}}
  {\ooalign{\raisebox{.14\height}{\scalebox{.7}{$\textstyle\sum$}}\cr\hidewidth$\textstyle\int$\hidewidth\cr}}
  {\ooalign{\raisebox{.2\height}{\scalebox{.6}{$\scriptstyle\sum$}}\cr$\scriptstyle\int$\cr}}
  {\ooalign{\raisebox{.2\height}{\scalebox{.6}{$\scriptstyle\sum$}}\cr$\scriptstyle\int$\cr}}
}

\begin{document}

\title{Semiclassical instanton formulation of Marcus--Levich--Jortner theory}

\author{Eric R. Heller}
\email{eric.heller@phys.chem.ethz.ch}
\author{Jeremy O. Richardson}
\email{jeremy.richardson@phys.chem.ethz.ch}
\affiliation{Laboratory of Physical Chemistry, ETH Z\"urich, 8093 Z\"urich, Switzerland}

\date{\today}

\begin{abstract}

Marcus--Levich--Jortner (MLJ) theory is one of the most commonly used methods
for including nuclear quantum effects into the calculation of electron-transfer rates
and for interpreting experimental data.
It divides the molecular problem into %
a subsystem treated quantum-mechanically by Fermi's golden rule
and a solvent bath treated by classical Marcus theory.
As an extension of this idea, we here present a ``reduced'' semiclassical instanton theory,
which is a multiscale %
method for 
simulating quantum tunnelling of the subsystem in molecular detail
in the presence of a harmonic bath.
We demonstrate that instanton theory is typically significantly more accurate than the cumulant expansion or the semiclassical Franck--Condon sum, which can give orders-of-magnitude errors and in general do not obey detailed balance.
As opposed to MLJ theory, which is based on wavefunctions, instanton theory is based on path integrals
and thus does not require solutions of the Schr\"odinger equation,
nor even global knowledge of the ground- and excited-state potentials within the subsystem.
It can thus be efficiently applied to complex, anharmonic multidimensional subsystems
without making further approximations.
In addition to predicting 
accurate rates, instanton theory gives a high level of insight into the reaction mechanism
by locating the dominant tunnelling pathway %
as well as providing information on the reactant and product vibrational states involved in the reaction
and the activation energy in the bath similarly to what would be found with MLJ theory.

\end{abstract}

\maketitle

\section{Introduction}

The theoretical study of electron-transfer is of essential importance and relevance not only because these reactions %
are a key step in many chemical and biological processes
but also because the methods developed to deal with them can be applied in many other scenarios %
ranging far beyond their original scope.
This follows from the fact that electron-transfer reactions are just one example of the more general set of curve-crossing problems. %
Hence, contributions to the understanding of electron-transfer reactions have been made %
with various motivations
including
electrochemistry, molecular spectroscopy, %
polaron transport as well as more general atom-transfer reactions,
which led to different ways of tackling the problem
from classical dielectric continuum theory to a full quantum molecular picture. \cite{UlstrupBook,KuznetsovBook}

Inspired by earlier work,\cite{Libby1952} Marcus based his theory of electron transfer, for which he later won the Nobel prize in 1992,\cite{Marcus1993review} first in terms of a dielectric solvent continuum \cite{Marcus1956ET} and later on a classical statistical mechanical description of the solvent.\cite{Marcus1960ET} 
To this day, Marcus theory is probably the most commonly applied approach for the description of electron-transfer reactions and initiated tremendous development involving electron and hole transfer between atoms, molecules or even proteins, in the condensed phase as well as at interfaces.\cite{Marcus1964review,Marcus1982review}
Hence, his findings had and still have an enormous impact on
a multitude of scientific disciplines comprising solution chemistry, solid-state physics as well as biological processes.\cite{Marcus1999acp}

One of the essential insights from Marcus' classical theory was the prediction of the so called ``inverted regime'',\cite{Marcus1960ET} the existence of which was later confirmed by experiment,\cite{Miller1984inverted} where the rate decreases as the thermodynamic driving force grows larger than the reorganization energy.
Soon, however, it was realized by theory and experiment that the neglect of nuclear quantum effects in Marcus theory can lead to dramatic errors of several orders of magnitude in the rate, especially in the inverted regime.\cite{Siders1981inverted,Efrima1976,Marcus1985review}

Based on the connection to spectroscopy and solid-state nonradiative processes Levich and coworkers
put the theory onto a rigorous quantum-mechanical basis and introduced a quantum statistical mechanical description of outer sphere electron transfer. \cite{Levich1966ET}
This was done by employing Fermi's golden rule\cite{Fermi1974} formula for the quantum transition rate, which is obtained as the nonadiabatic (weak-coupling) limit from perturbation theory.\cite{Dirac1927,Wentzel_1927}
However, because outer-sphere electron-transfer is typically dominated by the low-frequency solvent modes, the resulting quantum effects are rather small. %

Several years later, 
crucial advancements
were made in particular by %
Jortner and coworkers
by explicitly taking the reorganization of the inner sphere into account.\cite{Levich1970rate,Kestner1974,Ulstrup1975,Jortner1976,Jortner1988,UlstrupBook,KuznetsovBook}
As opposed to the solvent, the inner sphere often exhibits intra- and intermolecular rearrangements associated with high-frequency vibrational modes which are therefore subject to substantial quantum effects.
Hence, they treated the inner sphere quantum-mechanically using Fermi's golden rule while keeping the classical approximation for the solvent bath. The resulting Marcus--Levich--Jortner (MLJ) theory constituted a considerable progress for the whole field, as it was the first rigorously derived method able to describe nuclear quantum effects in electron-transfer reactions which was valid throughout virtually the whole temperature range.\cite{Bixon1999acp} Thus, the method poses a vital step towards the goal of establishing a unified description of electron transfer in the various fields mentioned above across diverse time and temperature scales.\cite{Marcus1982review,Closs1988,Barbara1992}

MLJ theory is broadly applied to the prediction and explanation of charge-carrier mobilities\cite{Duan2012} often with the objective to give a guideline for the synthetic study and reasonable design of high-performance
semiconductors that can be applied in organic photovoltaics.\cite{Pourtois2002,Lemaur2005,Geng2011,*Geng2011a,Liu2016}
The application to large systems can be facilitated by using the theory in conjunction with density-functional theory.\cite{Chaudhuri2017}
Furthermore it can be applied in the study of
molecular junctions,\cite{Thomas2019} 
photonics,\cite{Lanzani2012} polaritons\cite{GonzalezAngulo2019} and
polarons\cite{Asadi2013} as well as for the description of spin transitions and phosphorescence.\cite{Harvey2007,Veldman2008,Marian2011,Wozna2015,Samanta2017}
Besides these technological disciplines, it is also frequently applied for the understanding of complex chemistry,\cite{Walker1991} electron transfer in supermolecules\cite{Bixon1993} and 
biochemistry,\cite{Chance1979,Lee2000,Giese2002} charge transfer in DNA\cite{Jortner1998,*Bixon1999,Giese2000,Bixon2002} and
photosynthesis.\cite{Joran1987,Bixon1995}
As tunnelling is especially prevalent in the Marcus inverted regime, \cite{Bixon1991ET}
MLJ theory is of particular interest in the study of molecular electron-transfer reactions 
which are strongly exothermic
\cite{Liang1990,Akesson1991,*Akesson1992,Moser1993,Bregnhoj2016}
or which are initiated by photoexcitation.\cite{Chen1991,Rosspeintner2013photochemistry}

The MLJ description of quantum tunnelling, which will be extensively discussed in the later sections, is understood by shifting the Marcus parabolas (free energy along the bath coordinates) by the quantized energy levels of the inner sphere. The rate therefore consists of contributions from multiple vibrational channels weighted according to a thermal distribution.\cite{Barbara1996} This interpretation appears quite different from the standard picture of tunnelling in which a particle penetrates a potential energy barrier with an energy smaller than the barrier height.
The main disadvantage of the approach is that it requires wavefunction solutions of the
time-independent Schr\"odinger equation
in order to compute the energy levels and Franck--Condon overlaps, which severely limits its usefulness for the description of realistic, anharmonic and multidimensional systems.

A number of alternative methods which can be used for the study of electron-transfer reactions
and other golden-rule processes
\cite{Wolynes1987nonadiabatic,Bader1990golden,Cao1995nonadiabatic,*Cao1997nonadiabatic,*Schwieters1998diabatic,Kretchmer2013ET,Shushkov2013instanton,Menzeleev2014kinetic,Tao2019RPSH,Lawrence2019isoRPMD,Lawrence2018Wolynes,Lawrence2019ET,Lawrence2020rates,GRQTST,*GRQTST2,*Fe2Fe3,Huo2013PLDM,Duke2016Faraday,vibronic,Shi2004goldenrule,*Sun2016goldenrule,Karsten2018vibronic}
are based on Feynman's path-integral description of quantum mechanics\cite{Feynman1948PI,Feynman} rather than on wave mechanics.
From this set, semiclassical golden-rule instanton theory\cite{GoldenGreens,AsymSysBath} in particular bears multiple appealing features. Without any prior knowledge about the analytic shape of the potential, it locates the ``instanton''
in the full-dimensional configuration space of the system, which can be thought of as the optimal tunneling pathway,\cite{Miller1975semiclassical,Perspective,AdiabaticGreens,QInst} and therefore provides direct insight into the reaction mechanism.
Furthermore the method was recently extended towards the Marcus inverted regime,\cite{GoldenInverted} which otherwise typically poses a problem for imaginary-time path-integral approaches,\cite{nonoscillatory}
although some extrapolation techniques have been used successfully to avoid this problem in other methods. \cite{Lawrence2018Wolynes}
By employing a ring-polymer discretization to the paths,\cite{GoldenRPI} the instanton method is able to simulate tunnelling in multidimensional, anharmonic systems in a computationally efficient way and is ideally suited for calculations in conjunction with high-level electronic structure methods just as in the standard adiabatic formulation of the theory.\cite{Perspective,InstReview,RPInst,Andersson2009Hmethane,Rommel2012enzyme,hexamerprism,porphycene,dimersurf,GPR,*Muonium}
Golden-rule instanton theory constitutes a semiclassical path-integral formulation of Fermi's golden rule and hence has the potential to be applied in a multitude of different fields, just as the golden rule itself.

One of the great strengths of instanton theory is its full-dimensional formulation of tunnelling such that it does not rely on an \emph{a priori} choice of the reaction coordinate. However, for many relevant reactions, especially in the condensed phase, 
even if one does not know the exact tunnelling path, 
one already has a good idea of which part of the system under investigation has to be considered explicitly and which part can be accounted for on a coarser level. 
It is this same separation into an inner and outer sphere which was the cornerstone of MLJ theory. 
Therefore in this paper, the formalism for a ``reduced'' semiclassical golden-rule instanton theory will be laid out, which describes tunnelling within the modes of the inner sphere under the implicit influence of either a classical or quantum harmonic bath. The presence of the bath affects the equations of motion of the inner sphere and renders the resulting reduced instanton non-energy-conserving due to energy exchange between inner and outer sphere. This is analogous to the reduced density matrix formalism employed in the study of open quantum systems. %
The resulting instanton picture preserves the convenient interpretation of quantum tunnelling as a particle travelling in the classically forbidden region below the barrier. %

Although the formalisms seem at first glance rather different, 
we will draw a connection between the MLJ and instanton theories
by deriving them both from a common expression.
In doing so, it will be shown clearly that
the instanton approximation is fundamentally different from other approximations such as
the broadly applied cumulant expansion method\cite{KuehnBook} 
and the semiclassical Franck--Condon sum. \cite{Siders1981quantum,Siders1981inverted}
Numerical results demonstrate that instanton theory is very accurate over a range of systems including anharmonic modes where these alternative approximations break down.

Some rate theories have the advantage that they are based on expressions which are simple enough that one can easily
see their dependence on certain parameters and thus
gain insight into the behaviour of different systems. \cite{Tang1994,KuznetsovBook}
We will argue that %
instanton theory allows for a well-balanced combination of easily attainable insights,
as well providing a realistic molecular simulation. %
Even when applied to complex anharmonic multidimensional potentials,
the method uniquely identifies an optimal tunnelling pathway which provides
a simple one-dimensional picture of the reaction, %
highlighting which modes are involved in the tunnelling.
In addition to this, the instanton can be analysed to obtain
information on the energies of the initial and final states of the system before and after the electron-transfer event, similar to what is computed in MLJ theory as we will show.

\section{Golden-rule rate}
\label{sec:general}

The total Hamiltonian which describes electron transfer between 
a reactant $\ket{0}$ and product $\ket{1}$ electronic state
is defined by\cite{ChandlerET}
\begin{align} \label{H}
	\op{H} &= \op{H}_0 \ketbra{0}{0} + (\op{H}_1 - \varepsilon ) \ketbra{1}{1} + \Delta \big( \ketbra{0}{1} + \ketbra{1}{0} \big) ,
\end{align}
where the electronic interaction between the states is given by the nonadiabatic coupling  $\Delta$.
Throughout this work the electronic coupling is taken to be constant, but the generalization to position-dependent couplings is fairly straight forward. %
\cite{GoldenGreens}$^,$%
\footnote{For intermolecular electron transfer one can either use steepest descent or numerically integrate the results over a range of donor--acceptor distances\cite{UlstrupBook}}
Furthermore the coupling is assumed to be very weak such that the rates occur in the golden-rule limit, i.e.\ $\Delta\rightarrow0$, %
which is typically the case in electron-transfer reactions. \cite{ChandlerET}
A driving force, $\varepsilon$, has been included explicitly in the total Hamiltonian,
which could describe an internal energy bias
or the effect of an external field.
It is kept separate here for clarity but it could of course be simply absorbed into the definition of $\op{H}_1$.

We are interested in studying problems which can be subdivided into an inner sphere, whose molecular structural characteristics will be explicitly taken into account,
and an outer sphere, which typically includes the solvent degrees of freedom and will be treated as an effective
harmonic environment %
characterized by its spectral density.
In the language of open quantum systems, these are called subsystem and bath and are here taken to be uncoupled to each other, %
\footnote{Note that it would be possible to extend the reduced instanton theory of this paper to include linear coupling to the harmonic bath.}
although there is of course still some coupling through the nonadiabatic terms in \eqn{H}.
Hence, the full nuclear Hamiltonian for electronic state $\ket{n}$ can be written as
\begin{align}
	\label{equ:H_div}
	\op{H}_n &= \op{H}^{\text{s}}_n + \op{H}^{\text{b}}_n ,
\end{align}
where
\begin{subequations}
\label{equ:H_sb}
\begin{align}
    \label{equ:h_system}
    \op{H}_n^\text{s} &= \sum_{k=1}^{d} \frac{\op{p}_k^2}{2m} + V_n^\text{s}(\op{\mathbf{q}}) ,
    \\
    \label{equ:h_bath}
    \op{H}_n^\text{b} &= \sum_{j=1}^{D} \frac{\op{P}_j^2}{2M} + V_n^\text{b}(\op{\mathbf{Q}}) .
\end{align}
\end{subequations}
The subsystem Hamiltonians, $\op{H}^{\text{s}}_n$, only depend on the coordinates $\mathbf{q}=(q_1,\dots,q_{d})$ and their conjugate momenta $\mathbf{p}=(p_1,\dots,p_{d})$, while the bath Hamiltonians, $\op{H}^{\text{b}}_n$, are solely a function of the coordinates $\mathbf{Q}=(Q_1,\dots,Q_{D})$ and momenta $\mathbf{P}=(P_1,\dots,P_{D})$.
Without loss of generality, these degrees of freedom have been mass-weighted such that all subsystem modes are associated with the same mass, $m$, and likewise all bath modes with mass $M$.
The harmonic approximation for the bath will be employed:\cite{UlstrupBook}
\begin{equation}
    V_{0/1}^\text{b}(\mathbf{Q}) = \sum_{j=1}^{D} \thalf M \Omega_j^2 (Q_j \pm \zeta_j)^2 ,
    \label{equ:pes_bath}
\end{equation}
where the plus sign corresponds to the reactant state and minus sign to the product state.
The bath Hamiltonians thus combine in \eqn{H} to describe a spin-boson model,\cite{Weiss} %
defined by the associated frequencies $\{\Omega_j\}$ and displacements $\{\zeta_j\}$, which 
can be selected such that they represent an appropriate spectral density.
This spin-boson model is complemented by the subsystem modes, 
whose potential-energy surfaces will be kept general for the derivations in this work
such that they can, in principle, provide a realistic description of an anharmonic molecule.

The full system is prepared as a thermal equilibrium ensemble in the reactant state
with inverse temperature $\beta = 1/k_{\text{B}}T$ and partition function $Z_0 = \Tr \big[ \eu{-\beta \op{H}_0} \big]$.
The quantum-mechanical rate expression
for a reaction from the reactant to the product electronic state
in the golden-rule regime %
can be derived from a perturbation expansion to lowest order in the nonadiabatic coupling $\Delta$ between the two electronic states of
an integral over the flux correlation function\cite{ChandlerET}$^,$%
\footnote{In this limit the flux correlation function splits into two similar terms which, although they are not equal, each integrates to the same result.\cite{nonoscillatory}
Therefore only one of these terms is required\cite{Wolynes1987nonadiabatic}
which simplifies the derivation relative to the equivalent adiabatic approach%
\cite{AdiabaticGreens,InstReview,QInst}}
to give 
\begin{align}
    k(\varepsilon) Z_0
    &= \frac{\Delta^2}{\hbar^2} \int_{-\infty}^{\infty} \Tr\big[\eu{-(\beta\hbar-\tau - \iu t)\op{H}_0 /\hbar} \, \eu{-(\tau + \iu t) (\op{H}_1 - \varepsilon) /\hbar} \big] \, \rmd t .
    \label{equ:k_et}%
\end{align}%
The flux-correlation function is an analytic function of time, and hence, the rate is independent of the imaginary-time parameter $\tau$,\cite{Miller1983rate}
although it has been included explicitly as it will play a pivotal role in the semiclassical approximations taken later on.

Due to separability of the subsystem and bath, a quantum trace can be taken independently over the respective contributions. The reactant partition function thus factorizes according to $Z_0 = Z_0^{\text{s}}Z_0^{\text{b}}$ into 
a subsystem part, $Z^{\text{s}}_0 = \Tr_{\text{s}}\big[ \mathrm{e}^{-\beta \hat{H}^{\text{s}}_0} \big]$, and a bath part, $Z^{\text{b}}_0 = \Tr_{\text{b}}\big[ \mathrm{e}^{-\beta \hat{H}^{\text{b}}_0} \big]$. 
The correlation function likewise splits into product of subsystem and bath parts.

The rate can thus be rewritten using the convolution theorem of Fourier transforms:\cite{Ulstrup1975}
\begin{equation}
    \label{equ:conv}
    k(\varepsilon) = \frac{\Delta^2}{2\pi\hbar^3} \int I^\text{s}(v) I^\text{b}(\varepsilon - v) \, \rmd v ,
\end{equation}
where the subsystem and bath lineshape functions are
\begin{subequations} \label{equ:l_gen}
\begin{alignat}{2}
    \notag
    I^\text{s}(v) &= \left(Z_0^\text{s}\right)^{-1} \! &&\int_{-\infty}^\infty
    \Tr_{\text{s}}\big[\eu{-(\beta\hbar-\tau - \iu t )\op{H}_0^\text{s}/\hbar} \\ 
    \label{equ:lss_gen}
    && &\times \eu{-(\tau + \iu t )(\op{H}_1^\text{s} - v)/\hbar}\big]
    \, \rmd t ,\\ 
    \notag
    I^\text{b}(\varepsilon - v) &= \left(Z_0^\text{b}\right)^{-1} \! &&\int_{-\infty}^\infty
    \Tr_{\text{b}}\big[\eu{-(\beta\hbar-\tau - \iu t )\op{H}_0^\text{b}/\hbar} \\ 
    \label{equ:lsb_gen}
    && &\times \eu{-(\tau + \iu t )(\op{H}_1^\text{b} - \varepsilon + v)/\hbar}\big]
    \, \rmd t .
\end{alignat}
\end{subequations}
The equivalence to \eqn{equ:k_et} can easily be checked by 
substituting \eqs{equ:l_gen} into \eqn{equ:conv} after renaming the integration variable $t$ in the two cases to $t^\text{s}$ or $t^\text{b}$
and using
$\int \eu{\iu(t^\text{s}-t^\text{b})v/\hbar} \, \rmd v = 2\pi\hbar\,\delta(t^\text{s}-t^\text{b})$.

The lineshape function of the subsystem expanded simultaneously in the position and eigenstate bases can be written as
\begin{widetext}
\begin{align}
    \label{equ:lss}
    I^\text{s}(v) = \left(Z_0^\text{s}\right)^{-1} \iiint_{-\infty}^{\infty} \sum_{\mu} \sum_{\nu} 
    \eu{-(\beta\hbar-\tau-\iu t) E_0^{\mu}/\hbar} \,
    \eu{-(\tau+\iu t) (E_1^{\nu} - v)/\hbar} \,
    \braket{\mathbf{q}'|\psi_0^{\mu}}\braket{\psi_0^{\mu}|\mathbf{q}''} \braket{\mathbf{q}''|\psi_1^{\nu}}\braket{\psi_1^{\nu}|\mathbf{q}'}
    \, \rmd \mathbf{q}' \rmd \mathbf{q}'' \rmd t ,
\end{align}
\end{widetext}
where  $E_0^{\mu}$ and $E_1^{\nu}$ are the internal energy levels and $\psi_0^\mu$ and $\psi_1^\nu$ are the corresponding wavefunctions of reactants and products, respectively.
Because of the global harmonic approximation of the bath,
the well-known result for the spin-boson model\cite{UlstrupBook,Weiss} can be used to cast \eqn{equ:lsb_gen} into  
\begin{equation}
    \label{equ:lsb}
    I^\text{b}(\varepsilon - v) = \int \eu{-\Phi(\tau+\iu t )/\hbar - (\tau+\iu t ) (v - \varepsilon)/\hbar} \, \rmd t ,
\end{equation}
where the effective action of the bath is defined by %
\begin{equation}
    \label{equ:Sspinboson}
    \Phi(\tau) = \sum_{j=1}^{D} 2M\Omega_{j} \zeta_{j}^2 \left[ \frac{1 - \cosh{\Omega_{j}\tau}}{\tanh{\thalf\beta\hbar\Omega_{j}}} + \sinh{\Omega_{j}\tau} \right] .
\end{equation}
Note that this is an analytic function of its argument and can therefore also be used to describe real-time dynamics in \eqn{equ:lsb}.
The coordinate dependence of the bath has been completely integrated out,
which is the reason why the effective action [\eqn{equ:Sspinboson}] only depends on time.
Assuming the spectral density of the bath is known, %
the time-integral of \eqn{equ:lsb} can be carried out (either by quadrature or by steepest descent) 
in order to account for quantum effects within the solvent. %
\cite{Bader1990golden,Song1993quantum2} 

In cases where the bath represents a polar solvent environment,
which typically comprises long-wavelength polarization modes, it is often justified to approximate the effective action [\eqn{equ:Sspinboson}] by its classical, low-frequency limit
where $|\Omega_j\tau| \ll 1$ and $\beta\hbar\Omega_j \ll 1$. \cite{UlstrupBook} 
In this case, the classical bath action is
\begin{align}
    \Phi_{\text{cl}}(\tau) &= \Lambda^{\text{b}} \left(\tau - \frac{\tau^2}{\beta\hbar}\right) , 
    \label{equ:SBclassS}
\end{align}
where the bath reorganization energy is given by %
\begin{equation}
   \Lambda^{\text{b}} = \sum_{j=1}^D 2M\Omega_j^2 \zeta_j^2 .
   \label{equ:reorganize}
\end{equation}
In these formulas,
$\tau/\beta\hbar$ plays the role of a ``symmetry factor'' as described in \Ref{KuznetsovBook}.

In order to include a quantum harmonic bath with \eqn{equ:Sspinboson}, knowledge of the bath spectral density is required to define $\{\Omega_j\}$ and $\{\zeta_j\}$.
On the other hand, a classical harmonic bath [\eqn{equ:SBclassS}] can be simpler to employ as it
is fully characterized by its reorganization energy $\Lambda^{\text{b}}$
and thus requires much less information.

The approach which we will follow in this paper
is to evaluate the subsystem and bath lineshape functions %
using different representations and approximations
to derive multiple methods for computing electron-transfer rates in the golden-rule regime.

For instance, if the relaxation of the inner sphere is assumed to play no role in the reaction under consideration, there is no subsystem contribution to the Hamiltonians in \eqn{equ:H_div}. The subsystem lineshape function \eqn{equ:lss_gen} therefore reduces 
to $I^\text{s}(v) = \int \eu{+(\tau+\iu t) v/\hbar} \, \rmd t
= 2\pi\hbar \, \delta(v)$.
Employing the classical approximation for the action [\eqn{equ:SBclassS}] in the bath lineshape function \eqn{equ:lsb}, plugging the lineshape functions into \eqn{equ:conv}
and performing the final time-integral analytically leads the famous Marcus rate equation\cite{Marcus1985review}
\begin{equation}
    k_{\mathrm{MT}}^\text{b}(\varepsilon) = \frac{\Delta^2}{\hbar} \sqrt{\frac{\pi\beta}{\Lambda^\text{b}}}\,\mathrm{e}^{-\beta(\Lambda^\text{b} - \varepsilon)^2/4\Lambda^\text{b}}.
    \label{equ:marcus}
\end{equation}
It describes electron-transfer reactions which do not involve significant rearrangements within the inner sphere (subsystem) %
and are therefore determined only by the conformational changes in the outer sphere (bath). %
As is well known, Marcus theory thus gives the correct classical limit of the rate in the case of a spin-boson model.\cite{Levich1959,Levich1966ET,UlstrupBook}

In an alternative and more powerful derivation of Marcus theory, the trace could have been evaluated over the bath degrees of freedom in \eqn{equ:lsb_gen} directly by a classical phase-space integral. \cite{Schmidt1973,nonoscillatory}
In fact we could treat the subsystem in the same way to obtain $k_\text{MT}(\varepsilon)$, a theory equivalent to \eqn{equ:marcus} but written in terms of the total reorganization energy, $\Lambda=\Lambda^\text{s} + \Lambda^\text{b}$.
This treatment allows for anharmonic potential-energy surfaces but reduces to give the same rate formula as long as the free-energy surfaces themselves are harmonic.
In this case, one should treat the driving force $\varepsilon$ as a free energy as it can also include entropic effects. \cite{Marcus1984semiclassical}

In many cases, however, the inner sphere undergoes significant conformational changes as well and can therefore not be ignored. Moreover, the molecules in the reaction center commonly exhibit high-frequency modes, which necessitates the explicit consideration of quantum effects (such as the existence of zero-point energy and possibility of tunnelling) within an anharmonic environment.
We can derive various methods to compute the subsystem contribution simply by carrying out the sums and integrals of \eqn{equ:lss} in different orders. %
Although all these approaches give identical results in their exact form, they provide different starting points for taking approximations.

\section{Marcus--Levich--Jortner theory}

The %
subdivision of the full nuclear Hamiltonians of each electronic state into independent subsystem and bath parts [\eqn{equ:H_div}]
is the foundation on which MLJ theory is grounded. \cite{Levich1970rate,Kestner1974}
In this approach one then treats the subsystem quantum mechanically and the bath classically.

\subsection{Formalism}

Here we rederive MLJ theory on the basis
of the formalism laid out in Sec~\ref{sec:general}.
Starting from \eqn{equ:lss}, we first take the integrals over positions and time and set $\tau$ to be zero. Consequentially one arrives at Fermi's golden-rule (FGR) formula\cite{Zwanzig} for the lineshape function of the subsystem\cite{Kestner1974,Ulstrup1975}
\begin{align}
    \label{equ:k_fgr}
    I_{\text{FGR}}^{\text{s}}(v)  &= \left( Z^\text{s}_0 \right)^{-1} \sum_{\mu} \mathrm{e}^{-\beta E_0^{\mu}} \sum_{\nu}
    |\theta_{\mu\nu}|^2
    \, \delta(E_1^{\nu} - E_0^{\mu} - v) ,
\end{align}
where $\theta_{\mu\nu} = \int \psi_0^{\mu}(\mathbf{q})^* \psi_1^{\nu}(\mathbf{q}) \, \rmd \mathbf{q}$ are the Franck--Condon factors %
and the subsystem contribution to the reactant partition function in the energy eigenbasis is $Z^\text{s}_0 = \sum_\mu \mathrm{e}^{-\beta E_0^{\mu}}$. 

Combining this wavefunction representation of the subsystem part with the bath lineshape function [\eqn{equ:lsb}] using the classical effective action [\eqn{equ:SBclassS}] and performing the final convolution integral in \eqn{equ:conv} leads directly to
the Marcus--Levich--Jortner electron-transfer rate theory in a system of two crossing potentials of arbitrary shape in a classical harmonic bath\cite{Ulstrup1975}
\begin{equation}
    k_{\text{MLJ}}(\varepsilon) = \sum_\mu \sum_\nu k_{\mu\nu}(\varepsilon) ,
     \label{equ:k_bj_full}
\end{equation}
with
\begin{multline}
    k_{\mu\nu}(\varepsilon) = \frac{\Delta^2}{\hbar} \sqrt{\frac{\pi\beta}{\Lambda^{\text{b}}}}
    \frac{\mathrm{e}^{-\beta E_0^{\mu}}}{Z^{\text{s}}_0}\\ \times
     |\theta_{\mu\nu}|^2\,
    \eu{-\beta(\Lambda^{\text{b}} - \varepsilon + E_1^\nu - E_0^\mu)^2/4\Lambda^{\text{b}}} ,
    \label{equ:k_bj_comp}
\end{multline}
which is the most general version of MLJ theory used in this work.
The total rate in \eqn{equ:k_bj_full} comprises contributions from all reactant and product vibrational channels.

This ``static'' formulation of electron transfer (i.e.\ time has been integrated out) results in a rate expression that requires knowledge of %
all internal states of the subsystem Hamiltonians, $\op{H}_n^\text{s}$.
For complex anharmonic molecules, this is not possible to compute without further approximations.
Thus, the most commonly employed form of the Marcus--Levich--Jortner theory takes the extra approximation that the subsystem potentials for the reactant and product are displaced one-dimensional harmonic oscillators with identical frequencies, $\omega$. 
Motivated by the fact that in many problems of physical interest the subsystem comprises very high-frequency modes, %
it is often appropriate to assume that the thermal energy is low compared with the energy spacing in these modes. %
Hence, only transitions from the ground vibrational reactant state with quantum number $\mu = 0$ have to be considered. The general expression for the one-dimensional overlap integral of two displaced harmonic oscillator wavefunctions can therefore be further simplified, because only the terms\cite{Jortner1988}
\begin{equation}
    |\theta_{0\nu}|^2 = \frac{A^{\nu}\, \eu{-A} }{\nu!} ,
\end{equation}
with $ A = \Lambda^{\text{s}}/\hbar\omega$, 
have to be taken into account. This results in the well known rate formula for a single quantum harmonic mode in the low-temperature limit\cite{Jortner1976}
 \begin{align}
    \label{equ:k_bj}
    {k}_{\text{MLJ}} (\varepsilon) =  \sum_{\nu=0}^\infty  k_{\nu}(\varepsilon) ,\qquad %
    \hbar\omega \gg k_{\text{B}}T,
\end{align}
where the rate into the product-state $\nu$ is
\begin{equation}
    \label{equ:sres}
    k_{\nu}(\varepsilon) = \frac{\Delta^2}{\hbar} \sqrt{\frac{\pi\beta}{\Lambda^{\text{b}}}}\,
    \frac{A^\nu \, \eu{-A}}{\nu!}  \, 
    \eu{-\beta(\Lambda^{\text{b}} - \varepsilon + \nu\hbar\omega)^2/4\Lambda^{\text{b}}} .
\end{equation}
This low-temperature rate therefore consists of contributions from multiple, parallel product vibrational channels with effective driving forces of $\varepsilon_{\nu} = \varepsilon - \nu\hbar\omega$.

As can be seen from the exponential ``activation'' part of \eqn{equ:sres}, the major contributions to the rate will typically involve the product vibrational states whose effective driving force $\varepsilon_{\nu}$ is approximately equal to the bath reorganization energy.
In cases where $\varepsilon < \Lambda^\text{b}$, this is not possible, and so then the $\nu=0$ product state is expected to dominate.
Thus, the dominant product vibrational state will depend on the thermodynamic driving force
and 
transitions to highly excited vibrational states are of particular importance for very exothermic reactions and hence especially in the inverted regime.

\subsection{Model example}
\label{subsubsec:models}
The simple model which we will use to illustrate MLJ theory is formed of two-dimensional displaced harmonic oscillators
with one mode treated quantum mechanically and the other classically.
The subsystem potentials are defined by
\begin{align}
    V^{\text{s}}_{0/1}(q) &=  
    \tfrac{1}{2} m \omega^2(q \pm \xi)^2 ,
\end{align}
where the reactant state is associated with the plus sign and the product state with the minus sign,
and because $d=1$, we drop the mode index.
Given the frequency and reorganization energy, the displacements are defined by $\xi = \sqrt{\Lambda^{\text{s}}/{2 m \omega^2}}$.
The bath potentials are defined according to a $D=1$ version of \eqn{equ:pes_bath} with $\zeta = \sqrt{{\Lambda^{\text{b}}}/{2 M \Omega^2}}$.
In particular we will apply the theory to two different models, defined by the parameters in Table~\ref{table:1}, one of which is in the normal and the other in the inverted regime.

\begin{table}[b]
\caption{Definition of the parameters used in the two harmonic models studied in this work. %
The rate is independent of 
the masses $m$ and $M$, which therefore do not have to be defined.
As is common practice, the value of the frequencies are defined by their related wavenumber.
}
\label{table:1}
\begin{ruledtabular}
\begin{tabular}{lcc} 
 & Inverted-regime & Normal-regime\\
 & model & model\\
 \hline
 $T$ (K) & $300$ & $300$\\ 
 $\Lambda^{\text{s}}$ (\si{\Calorie\per\mol}) & $25$ & $50$\\ 
 $\Lambda^{\text{b}}$ (\si{\Calorie\per\mol}) & $25$ & $50$\\ 
 $\varepsilon$ (\si{\Calorie\per\mol}) & $75$ & $25$\\ 
 $\omega$ (\si{\per\cm}) & $1000$ & $500$\\ 
 $\Omega$ (\si{\per\cm}) & $50$ & $50$\\
\end{tabular}
\end{ruledtabular}
\end{table}
\begin{figure}
    \centering
    \includegraphics[width=0.48\textwidth]{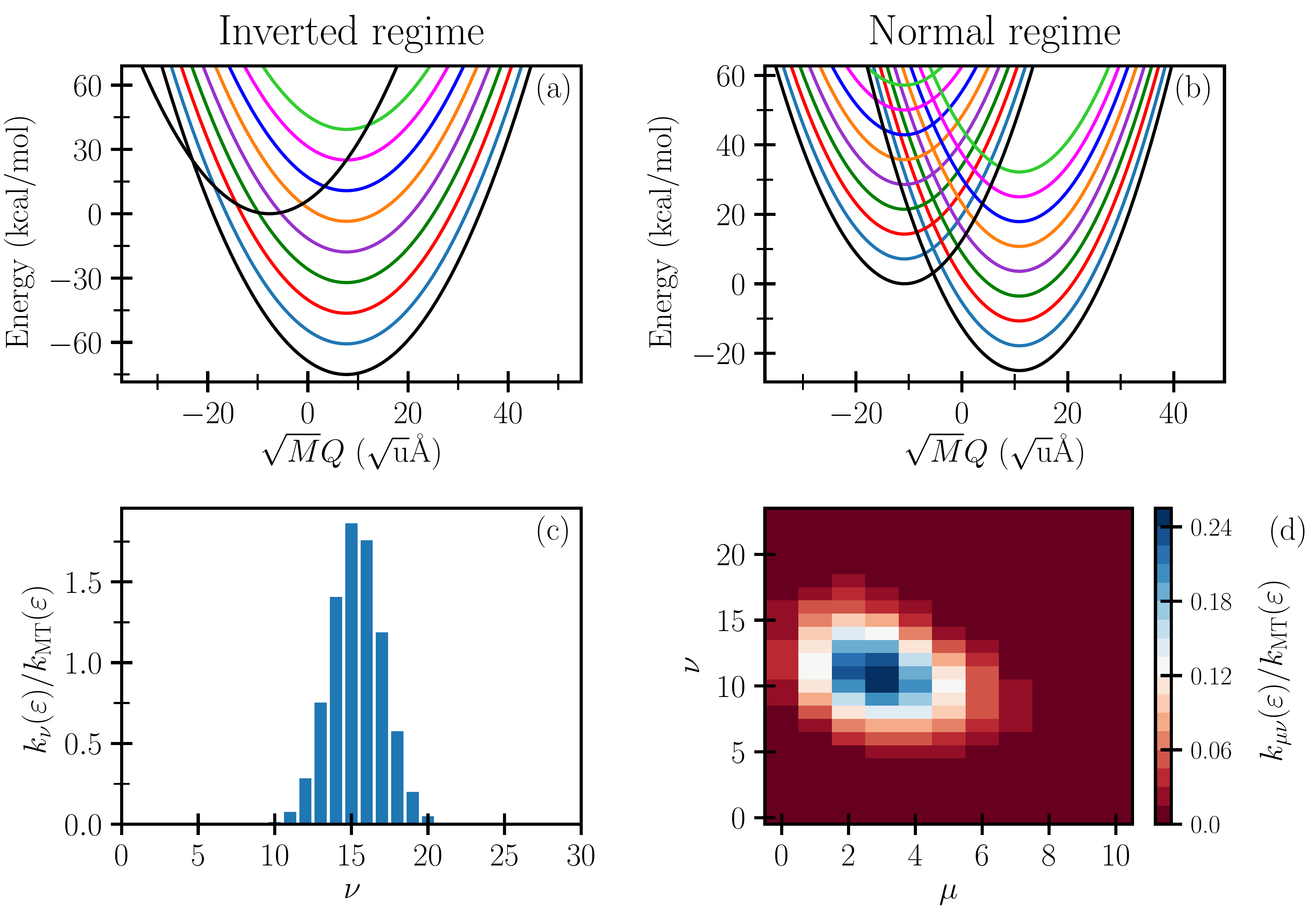}
    \caption{Plots to analyze the MLJ rate for the two harmonic models defined in Table~\ref{table:1}.
    (a,b) Plot of $V_0^\text{b}$ and $V_1^\text{b} - \varepsilon$ (black lines) as functions of the bath mode.
    The coloured lines are copies of the product potential shifted by the excitation energies of the quantized subsystem mode, and in (b) only, also shifted copies of the reactant potential.
    In each case, only every fifth state is shown.
    (c,d) State-resolved contributions to the MLJ rate relative to the corresponding Marcus theory rate for the full two-dimensional model. %
    For the inverted-regime model the reactant is almost always found in its vibrational ground state and therefore $\mu=0$. 
    } 
    \label{fig:pess}
\end{figure}

The parameters are chosen so as to illustrate 
two common scenarios.
Because the inner sphere typically comprises high-frequency vibrational modes, the low-temperature limit of the MLJ rate [\eqn{equ:k_bj}] can often be applied to a good approximation.\cite{Ulstrup1975,UlstrupBook} 
Hence, activated vibrational reaction channels only have to be considered for the product.
This case is exemplified by the inverted-regime model.
The situation is illustrated in Fig.~\ref{fig:pess}(a) which shows the product potential shifted by the vibrational energy gap $\nu\hbar\omega$.
In Fig.~\ref{fig:pess}(c) the reaction rate is broken down into contributions from the individual product channels, which are clearly centered around the dominant vibrational state $\nu=15$ and rapidly fall off on either side. 

Sometimes, however, a system
also requires the consideration of activated vibrational states of the reactant, which necessitates the use of the general expression \eqn{equ:k_bj_full}. 
As illustrated in Fig.~\ref{fig:pess}(d), this is the case for the normal-regime model, where excited vibrational states of both the reactant and product
make significant contributions to the rate. 
Thus, in Fig.~\ref{fig:pess}(b), not only the product but also the reactant potential is shifted by the vibrational energies.
The required Franck--Condon overlap integrals for a subsystem of two displaced harmonic oscillators can be computed with well-known analytic formulas.\cite{KuehnBook}
Fig.~\ref{fig:pess}(d) shows that the dominant contribution to the rate comes from the reaction channel from $\mu=3$ to $\nu=11$.

Perhaps even more important than the ability of MLJ theory to predict rates is that it provides this simple picture of the quantum nuclear effect on an electron-transfer reaction.
By viewing the reaction along the bath coordinates and shifting the potential-energy surfaces by the excitation energies of the subsystem, one obtains vibrational-state resolved contributions to the rate, which are centred around a dominant vibrational channel. This is a popular way of understanding reactions and has for instance been used to explain why rates in the inverted regime commonly flatten off instead of decreasing rapidly with driving force as predicted by classical Marcus theory [\eqn{equ:marcus}].\cite{Efrima1976,Miller1984inverted}
In the inverted regime, it can easily be seen from \eqn{equ:k_bj} and Figs.~\ref{fig:pess}(a) and (c) that the dominant contribution to the rate originates from the vibrational channel that approximately shifts the bath product potential to the activationless regime, where $\varepsilon_\nu \approx \Lambda^\text{b}$,\footnote{This simplistic picture would actually predict that $\nu=17$ rather than $\nu=15$ is the dominant product state.} %
and thus predicts a 
rate approximately independent of driving force. \cite{Efrima1974}

This analysis of the inverted-regime model is based on the simplification that the rate is fully determined by the exponential ``activation'' part. 
In reality, however, rates in the inverted regime are also affected by the Franck--Condon factors such that they do not actually become constant with driving force.
We find that the bath activation energy for the dominant vibrational transition in the inverted-regime model is
\SI{0.51}{\Calorie\per\mol}, which is almost activationless, but still not negligible relative to the thermal energy.
In the normal-regime model, where excited reactant states also play an essential role 
as illustrated in  Figs.~\ref{fig:pess}(b) and (d),
the full rate expression \eqn{equ:k_bj_full} is no longer dominated by an activationless channel at all.
We find a significant activation energy for the dominant vibrational channel of \SI{6.64}{\Calorie\per\mol},
which illustrates the compromise between minimization of the activation energy and maximization of the Franck--Condon overlaps that has to be made. 
This considerably complicates the interpretation of the MLJ rate formula even when the harmonic oscillator approximation is employed.

For more realistic systems described by multidimensional anharmonic potential-energy surfaces
the Franck--Condon factors are practically impossible to obtain
and the subsystem is thus commonly approximated by simple models for which these are known analytically.
This introduces unknown errors into the predicted rate, %
and it is to avoid this problem that we now turn to instanton theory.

\section{Reduced instanton theory} %
\label{subsec:inst_theory}

Inspired by the Marcus--Levich--Jortner approach we will derive a reduced instanton theory, where only the inner sphere is treated explicitly in molecular detail while the outer solvent shells are accounted for with the harmonic bath approximation.

\subsection{Formalism}
\label{subsec:rinst_form}

In order to derive the semiclassical golden-rule instanton rate expression, we start from the time-dependent correlation function formulation of the reaction rate in \eqn{equ:k_et}.
The trace can be split up into a subsystem and bath contribution, where the latter can, due to its harmonic nature, again be replaced with the well known solution for the spin-boson model in terms of the effective bath action $\Phi(\tau)$.  
In contrast to the MLJ approach, the trace in the subsystem coordinates will be expanded in the position basis, which leads the following expression for the rate:
\begin{multline}
    k \, Z_0^\text{s}   = \frac{\Delta^2}{\hbar^2} \iiint_{-\infty}^{\infty} K_0(\mathbf{q}',\mathbf{q}'',\beta\hbar - \tau - \mathrm{i}t ) \\
    \times K_1(\mathbf{q}'',\mathbf{q}',\tau + \mathrm{i}t) \,
    \eu{-\Phi(\tau+it)/\hbar + (\tau + \mathrm{i}t) \varepsilon/\hbar}
    \, \rmd \mathbf{q}' \, \rmd \mathbf{q}'' \,\rmd t .
    \label{equ:pi_class}
\end{multline}
Again in analogy to the dynamics of open quantum systems, $\eu{-\Phi(\tau+it)/\hbar}$ plays the role of an influence function.\cite{Feynman,Weiss}
The matrix elements of the quantum propagators,
\begin{align}
    \label{equ:qprop}
    K_n(\mathbf{q}_\text{i},\mathbf{q}_\text{f},\tau_n) &= \braket{\mathbf{q}_\text{f} | \mathrm{e}^{-\tau_n\hat{H}^{\text{s}}_n/\hbar} |\mathbf{q}_\text{i}} ,
\end{align}
describe the dynamics of the subsystem variables evolving 
according to the Hamiltonians $\hat{H}^{\text{s}}_n$ from the initial positions $\mathbf{q}_\text{i}$ to the respective final position $\mathbf{q}_\text{f}$
in imaginary time $\tau_n$.
The imaginary-time propagators are equivalent to quantum Boltzmann distributions
and it is this connection which allows instanton theory to approximate the thermal rate in a statistical way using imaginary-time dynamics.

If the imaginary-time propagators and spatial integrals were evaluated by path-integral Monte Carlo calculations and the remaining time integral taken by steepest descent, one would obtain a version of Wolynes theory where the bath is treated implicitly by the influence function.\cite{Wolynes1987nonadiabatic,Lawrence2019ET}
This, however, is not the purpose of this work as we wish to derive a semiclassical instanton formulation of the rate.

Instead we replace the quantum propagators by the corresponding van-Vleck propagators\cite{GutzwillerBook}  generalized for imaginary-time arguments\cite{Miller1971density,InstReview}
\begin{align}
    \label{equ:vVleck}
    K_n(\mathbf{q}_\text{i},\mathbf{q}_\text{f},\tau_n) \sim \sqrt{\frac{C_n}{(2\pi\hbar)^{d}}} \,\mathrm{e}^{- S_n/\hbar} ,
\end{align}
thus introducing a semiclassical approximation. The resulting expression is evaluated by locating the classical trajectory, $\mathbf{q}_n(u)$, travelling in imaginary time $u$,
which makes the Euclidean action of the subsystem, $S_n$, stationary.
The action for a path travelling from its initial position $\mathbf{q}_n(0) = \mathbf{q}_{\text{i}}$ to its final position $\mathbf{q}_n(\tau_n) = \mathbf{q}_{\text{f}}$ in imaginary time $\tau_n$
is defined as
\begin{align}
    S_n &\equiv S_n(\mathbf{q}_\text{i}, \mathbf{q}_\text{f}, \tau_n) = \int_{0}^{\tau_n} \left [ \thalf m \| \dot{\mathbf{q}}_n(u) \|^2 + V^{\text{s}}_n(\mathbf{q_n}(u)) \right] \rmd u ,
    \label{equ:action}
\end{align}
where $\dot{\mathbf{q}}_n(u) = \frac{\rmd \mathbf{q}_n}{\rmd u}$ is the imaginary-time velocity.
The prefactor of the semiclassical propagator is given by the determinant
\begin{equation}
C_n = \left| -\frac{\partial^2S_n}{\partial \mathbf{q}_\text{i} \partial \mathbf{q}_\text{f}} \right| \, .
\label{equ:C_n}
\end{equation}

By multiplying the two propagators in \eqn{equ:pi_class} together, we obtain the total action %
\begin{align}
    \label{equ:S_sys}
    S(\mathbf{q}',\mathbf{q}'',\tau) &= S_0(\mathbf{q}',\mathbf{q}'',\beta\hbar - \tau) + S_1(\mathbf{q}'',\mathbf{q}',\tau) ,    
\end{align}
as the sum of contributions from two trajectories, one of which travels on the reactant potential and the other on the product potential.
These trajectories join each other to form a continuous periodic pathway, called the instanton.
The imaginary times $\tau_n$ associated with the two paths are given by
$\tau_0 = \beta\hbar - \tau$ and $\tau_1 = \tau$.

Combining this result with the effective action of the bath according to \eqn{equ:pi_class},
the total effective action becomes
\begin{equation}
    \label{equ:S_tot}
    \mathcal{S}^\text{r}(\mathbf{q}',\mathbf{q}'',\tau) = S(\mathbf{q}',\mathbf{q}'',\tau) + \Phi(\tau) - \varepsilon \tau ,
\end{equation}
where one could employ either the effective quantum bath action from \eqn{equ:Sspinboson} or its classical limit \eqn{equ:SBclassS}.
The only effect of the bath is to thus alter the total action by adding an extra $\tau$-dependence alongside the driving force term.
However, as we will show, the simple addition of the bath action can lead to significant changes
for the instanton path and for our interpretation of the reaction mechanism.

In order to obtain the semiclassical instanton expression for the rate, 
the integrals over $\mathbf{q}'$ and $\mathbf{q}''$ as well as the time-integral will be carried out by steepest descent.
Therefore it is necessary first to study the path corresponding to the stationary point of the effective action, for which
$\pder{\mathcal{S}^\text{r}}{\mathbf{q}'}=\pder{\mathcal{S}^\text{r}}{\mathbf{q}''}=\pder{\mathcal{S}^\text{r}}{\tau}=0$. This path is our definition of the reduced instanton and
by analyzing the consequences of vanishing derivatives, we can understand its properties in the general case.

As in the standard golden-rule instanton formulation,\cite{GoldenGreens} the instanton pathway consists of two trajectories $\mathbf{q}_0(u)$ and $\mathbf{q}_1(u)$, which join smoothly into each other at the stationary or hopping point in the subsystem coordinate space $\mathbf{q}'=\mathbf{q}''=\mathbf{q}^\ddagger$.
Because the $\mathbf{q}$-derivatives are not altered by the influence of the bath, the momentum, given by %
$\mathbf{p}'=-\pder{S_0}{\mathbf{q}'} = \pder{S_1}{\mathbf{q}'}$
(equivalent for double primes), is thus still conserved across the hopping point. In this sense, the instanton therefore remains a periodic orbit of imaginary time $\beta\hbar$ as in the standard theory.\cite{GoldenGreens}

However, a major difference occurs due to the bath's influence on the derivative with respect to $\tau$.
The subsystem energies of the two trajectories are given by
\begin{align}
    E_n^{\text{s}} = \frac{\partial S_n}{\partial \tau_n} .
    \label{equ:Einst}
\end{align}
Therefore, in the case without the presence of a bath, the condition at the stationary point is given by $\pder{S}{\tau} - \varepsilon = 0$, where $\pder{S}{\tau}= E^\text{s}_1 - E^\text{s}_0 \equiv \Delta E^\text{s}$.
Considering $\varepsilon$ as a contribution to the product energy as was done in \Ref{GoldenGreens}, this relationship implies that the reaction conserves energy, i.e.\ $E^\text{s}_0 = E^\text{s}_1-\varepsilon$.
The hopping point must therefore be located on the crossing seam where $V_0(\mathbf{q}) = V_1(\mathbf{q}) - \varepsilon$.

This no longer holds true once bath modes are added. Then the condition at the stationary point changes to
\begin{equation}
    \pder{\mathcal{S}^\text{r}}{\tau} = \pder{S}{\tau} + \pder{\Phi}{\tau}-\varepsilon = 0 .
\end{equation}
The presence of the bath will thus affect the stationary value of $\tau$
and hence the entire instanton path and the value of its action.
\footnote{The classical bath action $\Phi_\text{cl}(\tau)$ is maximal at $\tau=\beta\hbar/2$
and assuming that the reaction is not endothermic, the stationary value will always obey $\tau \le \beta\hbar/2$.
Thus the inclusion of a classical bath will increase the stationary value of $\tau$ towards this limit
and may even cause a reaction to change from the inverted regime to normal regime.
It is also clear that the contribution from $\Phi_\text{cl}$ will be positive in the normal regime, but negative in the inverted regime.
}
In particular, the %
energies of the two trajectories no longer match $E^\text{s}_0 \ne E^\text{s}_1 - \varepsilon$ in general.
Hence, the presence of the bath renders the reduced instanton non energy-conserving within the subsystem. However, the energy change in the subsystem is exactly compensated by an energy change of opposite sign in the bath given by $\Delta E^\text{b} = \pder{\Phi}{\tau}$ such that $ \Delta E^\text{s} + \Delta E^\text{b} - \varepsilon = 0$. 
This is in agreement with what one would expect from an open quantum system,
in which only the total combined energy of subsystem and bath is conserved but not the individual components.
In this theory one does not have direct access to the energies of the bath which would be needed to fully justify this interpretation for $\pder{\Phi}{\tau}$.
However, we will show that this definition is correct in Sec.~\ref{sec:insights}.

One consequence of the energy jump caused by the presence of the bath is that the hopping point, $\mathbf{q}^\ddag$, is not located on the crossing seam between the two subsystem potentials.
In fact, because the momenta of the two trajectories are equal at the hopping point,
the energy jump within the subsystem must correspond exactly to the potential energy difference, $\Delta E^\text{s} = \Delta V^\text{s}(\mathbf{q}^\ddag)$, where $\Delta V^\text{s}(\mathbf{q}) = V_1^\text{s}(\mathbf{q}) - V_0^\text{s}(\mathbf{q})$.

\begin{figure}
    \includegraphics[width=0.48\textwidth]{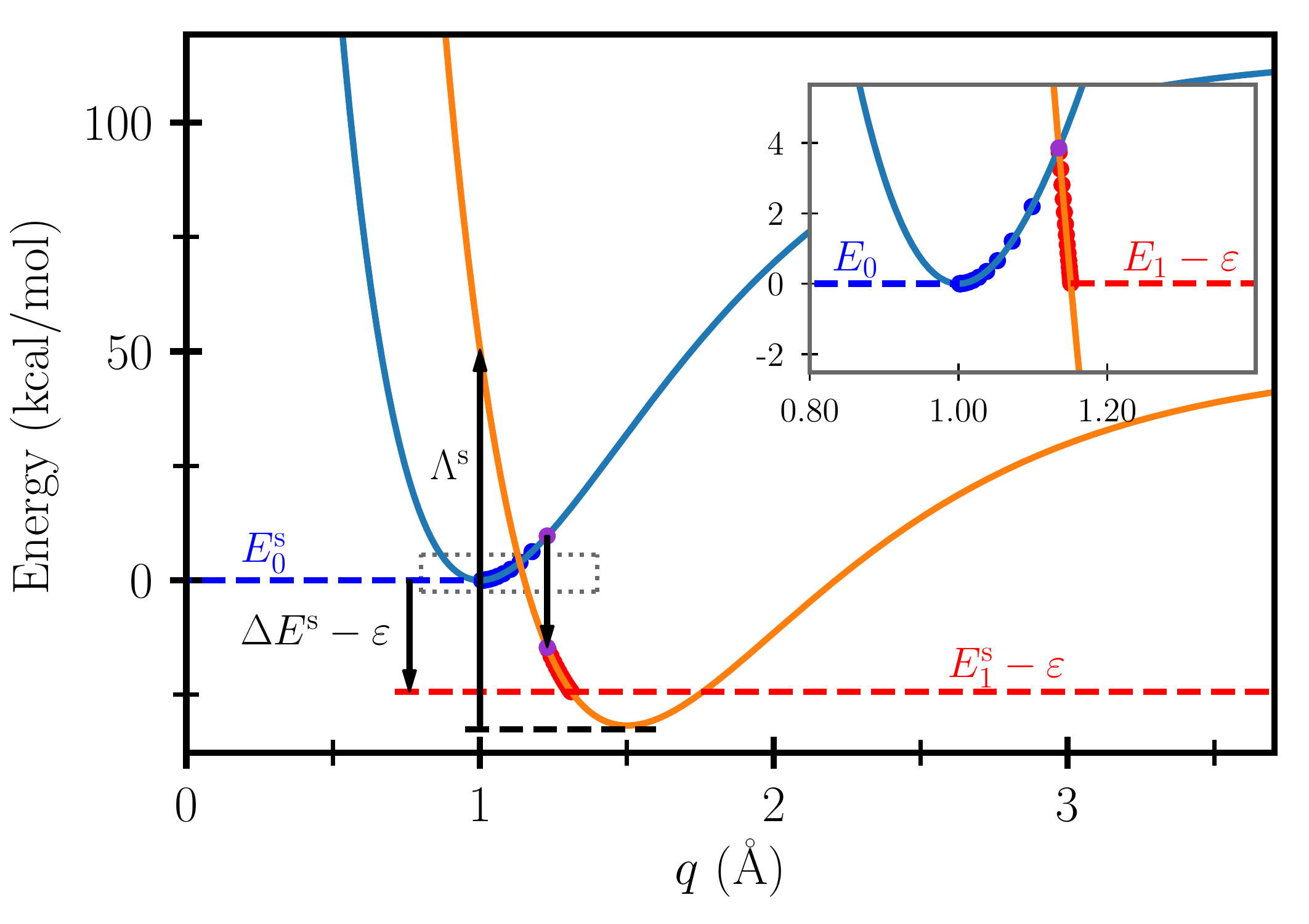}
    \caption{ 
    Potential curves $V^\text{s}_0(q)$ (blue solid lines) and $V^\text{s}_1(q) - \varepsilon$ (orange solid lines)
    of the subsystem as defined in \eqn{equ:anharmonic}
    with $\varepsilon=\Lambda/4 = \SI{31.7}{\Calorie\per\mol}$.
    The two trajectories of the reduced instanton are shown at discrete time steps by blue (reactant) and red (product) dots, and their 
    energies, $E_0^{\text{s}}$ and $E_1^{\text{s}} - \varepsilon$,
    are depicted by the blue and red dashed lines.
    These energies are separated by the energy gap $\Delta E^\text{s} - \varepsilon$, which
    is equal to the potential-energy gap at the hopping point ($q^\ddag$, purple dot).
    The inset shows the instanton for a model with the same subsystem but without the presence of a bath enlarged from the area framed by the grey dotted box. Here $\Delta E^\text{s} - \varepsilon = 0$ and therefore energy conservation is satisfied and the hopping point (purple dot) is located where the potentials cross.
    }
    \label{fig:Instcoupl}
\end{figure}

In Fig.~\ref{fig:Instcoupl}, we illustrate the reduced instanton pathway for the anharmonic model discussed in Sec.~\ref{subsubsec:morse}, as well as the 
the energies of the two trajectories
as defined by \eqn{equ:Einst}.

Note that the reactant or product energy is conserved along its respective trajectory and is thus identical to the potential at the turning point, which can be easily seen in the figure as the point with lowest potential along the path. The concept of a turning point in this context can be understood by the fact that dynamics in imaginary time are equivalent to real-time dynamics on the upside-down potential.\cite{Miller1971density}
At the turning point, the paths therefore bounce against the potential which they are travelling on.

Because the instanton orbit folds back on itself and is therefore not so easy to depict, it is worth describing it in a little more detail.
If we first follow the instanton pathway in Fig.~\ref{fig:Instcoupl} along the trajectory $q_0(u)$ starting at the (purple) hopping point on $V^\text{s}_0$ and with a certain amount of momentum pointing to the left, we find that the path descends towards the reactant state minimum, where it bounces against the potential and returns to where it started but with momentum now pointing to the right. So far this is equivalent to the instanton pathway in the standard formulation of the theory.\cite{GoldenGreens} Once the hopping point is reached, however, a sudden jump in potential energy occurs, which accompanies the transition into the product state. This is in stark contrast to the standard formulation, where both trajectories tunnel at the same energy, as shown in the inset of Fig.~\ref{fig:Instcoupl} for a case without a bath. 
After the state transition, we travel along the trajectory $q_1(u)$, which after reaching the turning point on $V^\text{s}_1$ also returns to the hopping point. A transition back to the initial hopping point on $V^\text{s}_0$ completes the periodic cycle. 

After the instanton pathway has been located, the integrals in \eqn{equ:pi_class}, where the propagators have been replaced with \eqn{equ:vVleck}, can be carried out by steepest-descent integration around the stationary point.
Thus we arrive at the reduced instanton expression for the golden-rule rate
\begin{equation}
    k_{\mathrm{rSCI}} (\varepsilon)\, Z_{0}^\text{s} = \sqrt{2\pi\hbar}\, \frac{\Delta^2}{\hbar^2}
    \sqrt{\frac{C_0 C_1}{C}} 
    \left( - \frac{\rmd^2 \mathcal{S}^\text{r}}{\rmd\tau^2} \right)^{-\frac{1}{2}} \mathrm{e}^{-\mathcal{S}^\text{r}/\hbar}
   \label{equ:krinst} ,
\end{equation}
where all quantities are evaluated at the stationary point of
$\mathcal{S}^\text{r}(\mathbf{q}',\mathbf{q}'',\tau)$ except the reactant partition function $Z_0^\text{s}$, which is treated by an equivalent steepest-descent approximation around the minimum of the reactant.\cite{InstReview}
The additional prefactor from the steepest descent integration in the subsystem positions evaluates to the determinant
\begin{equation}
C = \begin{vmatrix}
\frac{\partial^2S}{\partial \mathbf{q}' \partial \mathbf{q}'} & \frac{\partial^2S}{\partial \mathbf{q}' \partial \mathbf{q}''}\\
\frac{\partial^2S}{\partial \mathbf{q}'' \partial \mathbf{q}'}  & \frac{\partial^2S}{\partial \mathbf{q}'' \partial \mathbf{q}''}
\end{vmatrix} ,
\label{equ:c_matrix}
\end{equation}
where we have used the fact that derivatives of $\mathcal{S}^\text{r}$ with respect to the end points are equal to derivatives of $S$.
The bath therefore has no direct effect on $C$ but does explicitly appear in
$\frac{\rmd^2 \mathcal{S}^\text{r}}{\rmd\tau^2} = \der[2]{S}{\tau} + \der[2]{\Phi}{\tau}$
as well having an important effect on the instanton path itself as previously discussed.
Apart from these changes, the formula resembles 
the semiclassical golden-rule instanton rate expression derived in previous work\cite{GoldenGreens}
and gives identical results without needing to treat the harmonic bath explicitly.

Just as for previous golden-rule instanton calculations,\cite{AsymSysBath,GoldenInverted} a ring-polymer discretization scheme of the instanton pathway is employed in order to 
describe nonadiabatic reactions for multidimensional, anharmonic systems. 
 By adopting the ring-polymer formalism, the localization of the instanton path, which is defined as a stationary point of the action in \eqn{equ:S_tot} in the coordinate and $\tau$ variables together, reduces to a standard saddle-point search problem
which can be solved numerically with well-established optimization algorithms.
Algorithms for computing the necessary derivatives of the action
as well as detailed information about the optimization scheme can be found in \Ref{GoldenRPI}.

In our recent extension of the theory,\cite{GoldenInverted} we have shown that ring-polymer instanton theory can equivalently be utilized to compute electron-transfer rates in the Marcus inverted regime, where tunnelling effects commonly play a particularly important role.
The major difference in this regime is, that one of the two paths travels in negative imaginary time, which allows an analogy to the physics of antiparticles.\cite{Feynman1986antiparticles} 
In the computational realization, this difference manifests itself merely in a slight change of the optimization algorithm. Hence, whereas in the normal regime the instanton is a single-index saddle point of the ring-polymer action in the combined space of ring-polymer coordinates and imaginary time, in the inverted regime the instanton path corresponds to a higher-index saddle point of the ring-polymer action. The index of a saddle point here defines the number of negative eigenvalues in the second-derivative matrix of the ring-polymer hessian at this point.
But since we exactly know the index of the desired saddle-point, the instanton can be optimized with the same routines by using standard eigenvector-following schemes. We thus take uphill steps in the direction of eigenvectors corresponding to negative eigenvalues and standard down-hill steps in the direction of eigenvectors associated with positive eigenvalues. This methodology can be directly transferred to the reduced instanton picture without any additional complications and hence allows us to apply it to the normal and inverted regimes alike.

The advantage of the reduced instanton approach is that the optimization is confined to the inner sphere and $\tau$-coordinates only, whereas the only direct influence of the bath on the optimization procedure manifests itself in
an external field in the imaginary-time variable.
This reduces the computational costs of the simulation
and enables it to be applied within a multiscale modelling approach, where certain parts of a system are treated at higher levels of accuracy than others.

\subsection{Model example}
\label{subsubsec:morse}

We will employ the newly formulated reduced instanton method along with MLJ theory to compute reaction rates of an anharmonic subsystem of two bound Morse oscillators in a multidimensional harmonic bath.
The subsystem is defined by the potentials (depicted in Fig.~\ref{fig:Instcoupl})
\begin{equation}
    V^{\text{s}}_n(q) = D^{\text{e}}_n \left( 1 - \eu{-\alpha_n (q - \xi_n)} \right)^2,
    \label{equ:anharmonic}
\end{equation}
where $\alpha_0 = \SI{1.5}{\angstrom^{-1}}$ and $\alpha_1 = \SI{1.4}{\angstrom^{-1}}$ determine the length scales,
$\xi_0 = \SI{1.0}{\angstrom}$ and $\xi_1 = \SI{1.5}{\angstrom}$ are the equilibrium positions, and
$D^{\text{e}}_0 = \SI{115}{\Calorie\per\mol}$  and $D^{\text{e}}_1 = \SI{80}{\Calorie\per\mol}$ are the dissociation energies of reactants and products.
The (product) reorganization energy of this subsystem is therefore $\Lambda^{\text{s}} = V_1(q_\text{min}^{(0)})-V_1(q_\text{min}^{(1)}) = \SI{82.2}{\Calorie\per\mol}$, where $q_\text{min}^{(n)}$ is the minimum of $V_n(q)$. The reduced mass is chosen to be $m=\SI{1.10}{\amu}$.
The well frequencies of the two Morse oscillators obtained by harmonic analysis are $\omega_n = \alpha_n \sqrt{2 D_n^\text{e}/m}$, which
results in frequencies of $\omega_0 = \SI{2358}{\per\centi\meter}$ and $\omega_1 = \SI{1835}{\per\centi\meter}$ for the reactant and product well respectively.
The Schr\"odinger equation for the Morse oscillator can be solved analytically to give the bound-state energies
\begin{equation}
    E_0^{\mu} = \hbar \omega_0 (\mu + \thalf) - \hbar \omega_0 \chi_0 (\mu + \thalf)^2 
\end{equation}
and likewise for $E_1^\nu$,
where the  dimensionless anharmonicity parameters of the Morse oscillators are defined by
$\chi_n = \alpha_n^2 \hbar / 2 m \omega_n$
and in this case have values of $0.015$ and  $0.016$ for the reactant and product potential respectively. 

The bath is defined by the discretized spectral density
\begin{align}
    J(\Omega) = \frac{\pi}{2} \sum_{j=1}^{D} \frac{c_j^2}{M\Omega_j} \delta(\Omega - \Omega_j) ,
\end{align}
and the $D = 100$ bath modes were chosen according to \Ref{Wang1999mapping} as
\begin{align}
    \Omega_j = \frac{j^2}{D^2} \Omega_{\text{max}},
    \qquad
    j\in [1,D]
    .
\end{align}
The effective mass of the bath modes $M$ does not have to be specified, as %
the rate is independent of this choice.
The frequency spectrum is bounded from above by the maximum frequency $\Omega_{\text{max}} = \SI{3000}{\per\centi\meter}$ and thus has the density%
\footnote{We choose a spectral density with a well-defined maximum frequency to ensure that the time-integral over the correlation function converges even when $\tau<0$.}
\begin{align}
    \rho(\Omega) = \frac{D}{2 \sqrt{\Omega\Omega_{\text{max}}}} .
\end{align}
The couplings were chosen to emulate a Debye spectral density defined by
\begin{align}
    J_{\text{De}}(\Omega) = \frac{\eta\,\Omega_{\text{c}}\Omega}{\Omega^2 + \Omega_{\text{c}}^2} ,
\end{align}
with characteristic frequency $\Omega_{\text{c}} = \SI{500}{\per\centi\meter}$ and $\eta = \SI{25}{\Calorie\per\mol}$.
Hence the coupling constants $c_j$ are determined by the formula
\begin{align}
    c_j^2 = M \Omega_j \frac{2}{\pi} \frac{J_{\text{De}}(\Omega_j)}{\rho(\Omega_j)} ,
    \qquad
    j\in [1,D]
    ,
\end{align}
which are related to the shifts $\zeta_j$ in \eqn{equ:pes_bath} by $c_j = M \Omega^2_j \zeta_j$.
The reorganization energy of the bath is then obtained by \eqn{equ:reorganize},
which in our case results in $\Lambda^{\text{b}} = \SI{44.8}{\Calorie\per\mol}$. The total reorganization energy of the subsystem and bath combined is therefore given by $\Lambda = \Lambda^\text{s} + \Lambda^\text{b} = \SI{127.0}{\Calorie\per\mol}$.
The temperature for the rate calculations was chosen to be \SI{300}{\K}.

Similar models were studied with MLJ theory in \Ref{Sondergaard1976}. In order to make use of analytical formulas for the Franck--Condon factors, however, in that work the reactant's subsystem mode was assumed to be in the low-temperature limit. %
Although it only makes a minor difference,
here, we include as many reactant states as is necessary to 
converge the rate, and perform the Franck--Condon overlap integrals numerically. 
Where necessary, we take the continuum states of the Morse oscillator into account, by extending the MLJ formula given in \eqn{equ:k_bj_full} in the same way as explained for the exact quantum rate [\eqn{equ:k_exact}] in Appendix~\ref{app1}.

\begin{figure}
\includegraphics[width=0.48\textwidth]{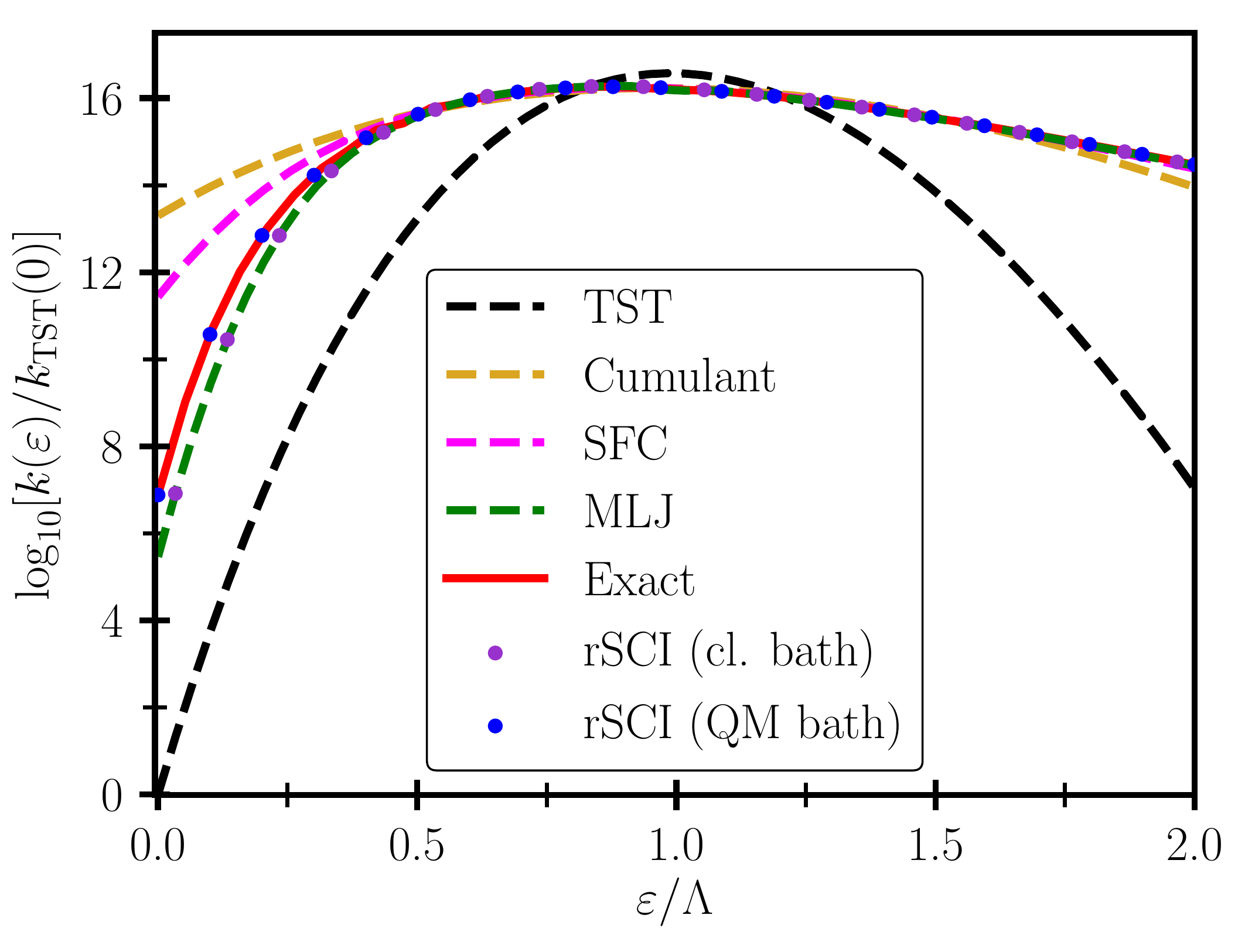}
    \caption{Rates calculated by various methods 
    for an anharmonic mode in conjunction with a harmonic bath
    are shown for different values of the driving force $\varepsilon$,
    including:
    the reduced semiclassical instanton [rSCI, \eqn{equ:krinst}] with either a quantum or classical bath; classical golden-rule transition-state theory [TST, \eqn{equ:TST}]; the second-order cumulant expansion [\eqn{equ:k_ce}]; the semiclassical Franck--Condon sum [SFC, \eqn{equ:sfc}]; Marcus--Levich--Jortner theory [MLJ, \eqn{equ:k_bj_full}] and exact quantum mechanics [\eqn{equ:k_exact}]. 
    In each case, the results are given relative to the classical golden-rule TST rate at $\varepsilon=0$. %
    }
    \label{fig:rates300}
\end{figure}

The reaction rates for this model system computed with various methods as a function of the %
driving force  $\varepsilon$ are presented in
Fig.~\ref{fig:rates300}.
The exact (Fermi's golden rule) and MLJ rate calculations are based on the knowledge of the analytic expressions for the energy levels and wavefunctions of the Morse oscillator (see Appendix~\ref{app1}). 
Thus, the only approximation made by MLJ is to treat the bath classically.
It can be seen from Fig.~\ref{fig:rates300} that instanton theory, which does not require knowledge of the eigenstates nor even global knowledge of the potential along the subsystem mode, is virtually identical to the exact result when employing a quantum bath with the effective action from \eqn{equ:Sspinboson}. 
This excellent agreement was expected from the results and analysis seen in previous instanton studies of electron transfer. \cite{AsymSysBath,GoldenInverted,GRQTST,GRQTST2}
When using a classical bath with the action given by \eqn{equ:SBclassS}, it is slightly less accurate,
although then very similar to MLJ theory as they both suffer from the 
assumption of a classical bath.

The classical golden-rule transition-state theory (TST) rate, outlined in Appendix \ref{app1}, constitutes the classical limit of the quantum rate and would reduce to Marcus theory in the case of a subsystem consisting of displaced harmonic oscillators. 
The deviation of the TST rate from the exact, MLJ and instanton rates underlines the importance of nuclear quantum effects, which causes the classical rate to differ from the exact results by more than seven orders of magnitude in some cases. The differences are most extreme for the largest driving forces in the inverted regime. 

For $\varepsilon>\Lambda$, the instanton analysis predicts a negative value of $\tau$,
which is a clear indication that the inverted regime has been reached, and it requires a subtly different ring-polymer optimization scheme. \cite{GoldenInverted}
Despite this, it is noteworthy that the observed turnover in the rate (i.e.\ the point at which the rates start to decrease with growing driving force) actually occurs at a slightly smaller driving force.
This is predicted correctly by all methods tested apart from classical golden-rule TST.
In instanton theory, this effect is caused by the prefactor in \eqn{equ:krinst} as the effective reduced action in the exponential has its minimum at $\varepsilon = \Lambda$.

The fact that in the inverted regime the MLJ, instanton and exact quantum rates almost coincide, reveals that practically all the quantum effects in this regime originate from the subsystem and not from the bath.
Although the quantum subsystem still plays a dominant role in the normal regime,
it is clear that there is also a small quantum effect from the bath,
which explains the source of the error of MLJ 
and likewise of rSCI theory when employing a classical bath. %

In Sec.~\ref{subsec:Inst_MLJ}, we will further elaborate on the relationship between rSCI and MLJ theory that is apparent from the results in this section.

\subsection{Comparison with alternative approximations}

In this subsection, we will compare the instanton approach
with two other approximate methods for including quantum effects into electron-transfer rates,
namely the cumulant expansion and the semiclassical Franck--Condon sum.
As well as discussing the accuracy of these various methods, we will also focus on the computational effort required for their calculation.

In Fig.~\ref{fig:rates300}, we present the rates obtained with the second-order cumulant expansion\cite{Kubo1962,Hashitsume1998,Loring1987} described in Appendix \ref{app2}.
To enable a direct comparison with MLJ theory, we employed a classical bath in its calculation,
although like with rSCI it would also be possible to use the effective action of a quantum bath.
This method is not only commonly used for the study of electron-transfer reactions in anharmonic systems,\cite{Sparpaglione1988,Hu1989,Borgis1991,Islampour1991,*Islampour1991a,*Islampour1993,Cho1995,Georgievskii1999,Soudackov2005,Renger2003} 
but also for the simulation of optical spectroscopy,\cite{Zhu2009} 
vibrational lineshapes\cite{KuehnBook} and the description of energy-transfer processes.\cite{Ma2015} In practice often further approximations are invoked to obtain analytical expressions for the rate\cite{Soudackov2015,*Soudackov2016} before the method can be applied to complex problems. %

The advantage of the cumulant expansion over an exact (FGR) or MLJ calculation
is that it does not require knowledge about the excited state's vibrational eigenstates, but only about its potential-energy surface. 
However, although the method is exact for displaced harmonic potentials,\cite{KuehnBook} the results in Fig.~\ref{fig:rates300} clearly demonstrate, that, as opposed to instanton theory, the rates obtained by the second-order cumulant expansion can differ significantly from the exact rates for the Morse oscillator model, with the worst case being at zero driving force in the normal regime.%
\footnote{We can understand this behaviour from a comparison with instanton theory, whose
value of $\tau$ gives a simple measure of the relative importance of the reactant and product dynamics in the calculation.  For $\varepsilon = \Lambda$, $\tau = 0$ which implies that all dynamics take place on the reactant, but at $\varepsilon = 0$, $\tau$ approaches $\beta\hbar/2$ (which is only strictly true for a symmetric system) so both are approximately equally important.
The cumulant expansion treats the reactant state on a higher level than the product state
and is thus expected to work best near the activationless regime ($\varepsilon = \Lambda$).
Hence, it is a good method for predicting optical lineshapes which are largest at this point, but not in general for rate calculations.}

Moreover the rate expression of the cumulant expansion does not satisfy the detailed balance relation for thermal rates, \cite{ChandlerGreen,Marcus1984semiclassical}
\begin{equation}
    k_{0 \rightarrow 1} \, Z^\text{s}_0 = 
    \eu{+\beta \varepsilon} \, k_{1 \rightarrow 0} \, Z^\text{s}_1 ,
    \label{equ:detbal}
\end{equation}
in anharmonic subsystems or even in a subsystem of two displaced harmonic oscillators of different frequency. Here, $k_{0 \rightarrow 1}$ and $ k_{1 \rightarrow 0}$ are the rate constants of the forward and backward reactions.
Note that this relation would normally be written with total partition functions, but here we have already used the fact that in our case $Z^\text{b}_0=Z^\text{b}_1$.
Detailed balance is however obeyed by Fermi's golden rule, MLJ theory, all forms of instanton theory and even classical golden-rule TST.

Another method that does not obey detailed balance for anharmonic subsystems 
is the ``semiclassical Franck--Condon sum'' (SFC). In fact, the rates computed within this approximation do not even fulfil detailed balance for a subsystem of two displaced harmonic oscillators of the same frequency if the driving force is different from zero.
Originally the method was developed to describe spectral line shapes of solids\cite{Lax1952,Curie1963} and later used to describe electron-transfer in biological systems.\cite{Hopfield1974,Chance1979} 
The derivation of the method for models with a harmonic bath as considered in this paper is outlined in Appendix~\ref{app3}.
In this case we treat the bath itself within the SFC approximation as this can be done with a closed-form expression.
Like the cumulant expansion, it requires knowledge of the vibrational eigenstates of the reactant but not of the product.
In accordance with the findings of Siders and Marcus,\cite{Siders1981quantum,Siders1981inverted}
it is accurate near the activationless regime, works fairly well in the inverted regime
and gives significant errors of more than four orders of magnitude in the normal regime.

Ultimately, the major intrinsic problem of the MLJ method is that it relies on 
knowledge of the wavefunctions of the reactant and product states and is therefore practically impossible to apply to complex multidimensional problems.
The cumulant expansion and SFC methods only go part of the way to improving this situation as they require wavefunctions only for the reactant state.  However, numerical integration over the coordinates would still require both potentials to be evaluated over a large grid.
Typically therefore at least the reactant potential is approximated by a low-dimensional harmonic oscillator,
which introduces an unknown additional error into the predicted rate.

In contrast, instanton optimizations only require information %
along the tunnelling pathway, which is located close to the hopping point and thus minimizes the computational effort.
This advantage of instanton theory over the wavefunction-based methods increases in significance with growing dimensionality of the subsystem. The reason for this is that the instanton pathway always remains one-dimensional, whereas the number of points needed to evaluate the potential-energy surfaces on a grid grows exponentially with subsystem size.
It can thus be applied in principle to complex systems without making extra approximations.

It is therefore worth noting that although our instanton approach as well as the SFC method and a number of other theories are labeled ``semiclassical'',
they clearly employ quite different approximations.
Not only is semiclassical instanton theory superior in accuracy, it is also applicable to more complex multidimensional anharmonic problems.

\section{Instanton formulation of MLJ theory}
\label{subsec:Inst_MLJ}

Although MLJ and reduced instanton theory can both be derived from \eqn{equ:k_et}, the resulting methods and rate formulas [\eqn{equ:k_bj_full} and \eqn{equ:krinst}] look rather distinct from each other
and thus lead to quite different interpretations of the reaction. 
Marcus--Levich--Jortner theory relies on the wavefunction picture of quantum mechanics and
computes the rate as a sum
over reactant and product states which will be dominated by one particular reaction channel as shown in Fig.~\ref{fig:pess}(d).
Instanton theory, on the other hand, is based on the path-integral formalism of quantum mechanics and
is dominated by a path which describes the mechanism during the electron-transfer event.

Another fundamental difference between MLJ theory and the rSCI approach presented in Sec.~\ref{subsec:inst_theory}
is that, in rSCI, the focus is shifted from the bath modes to the subsystem.
The standard MLJ picture as shown in Fig.~\ref{fig:pess}
interprets the reaction in terms of the activation energy in the bath and includes the effect of the subsystem through the shift that they give to the bath potentials.
The computation of the reduced instanton approach, however, is carried out directly 
in the subsystem modes under the influence of the bath.
This reflects more appropriately the computational effort put into the calculation of subsystem and bath, as typically the subsystem will be treated in much more detail or on a higher level of theory.

Both interpretations can be useful, but 
it is not immediately obvious that they can be reconciled,
although the common foundation in \eqn{equ:k_et}
suggests that both methods must be related. 
This idea is reinforced by the fact that the rates obtained for the double Morse oscillator model, shown in Fig.~\ref{fig:rates300}, are practically identical when both methods treat the bath classically. 
In the following we will show that a different derivation of the semiclassical instanton approximation
leads to an equivalent formulation but which can be used to give the same insights as MLJ theory.

\subsection{Formalism}
\label{subsec:iMLJ}

The objective of this section is to derive an instanton formulation of MLJ theory.
The bath is thus assumed to be classical and
for simplicity both subsystem and bath are kept one-dimensional here.
The formulas do, however, generalize straightforwardly to the multidimensional case. 

In order to show the relation with MLJ theory more closely, 
the convolution formula  [\eqn{equ:conv}] will again serve as the starting point. 
The expression for the lineshape function of the bath in \eqn{equ:lsb_gen}, will be evaluated by a classical phase-space integral, which is one dimensional in both the position and momentum coordinate. After carrying out the integrals in momentum and time, this results in the one-dimensional classical configuration-space integral
\begin{multline}
    I_\text{cl}^\text{b} (\varepsilon - v) = 2\pi\hbar \left( Z_0^{\text{b}} \right)^{-1} \sqrt{\frac{M}{2\pi\beta\hbar^2}}\\
    \times
    \int \eu{-\beta V_0^\text{b}} \delta(\Delta V^{\text{b}} - \varepsilon + v) \, \rmd Q ,
    \label{equ:Ibath}
\end{multline}
where 
the  Hamiltonians of the bath [\eqn{equ:h_bath}] have been replaced by their classical analogues and the independence with respect to $\tau$ appears naturally. In addition, we define the potential energy difference in the bath $\Delta V^{\text{b}} (Q) = V^{\text{b}}_1(Q) - V^{\text{b}}_0(Q)$, although we suppress the $Q$-dependence to avoid clutter. 
For the harmonic bath potential, $\Delta V^{\text{b}}=-2M\Omega^2\zeta Q=-\Lambda^\text{b} Q/\zeta$. The $Q$-integral could of course easily be carried out immediately to give the Marcus theory lineshape. However, In order to obtain a picture of the reaction from the point of view of the bath, we leave it for later.

Using this classical result for the bath lineshape function in \eqn{equ:conv} and performing the convolution integral leads the approximate rate formula
\begin{align}
    \label{equ:pi_class2}
    k(\varepsilon)
    &\approx \left( Z_0^{\text{b}} \right)^{-1} \frac{\Delta^2}{\hbar^2} \sqrt{\frac{M}{2\pi\beta\hbar^2}} 
     \int \tilde{I} (\varepsilon - \Delta V^{\text{b}}) 
     \, \rmd Q ,
\end{align}
where we define the subsystem lineshape function weighted by the bath thermal distribution
\begin{equation}
    \tilde{I}(\varepsilon-\Delta V^{\text{b}}) =  I^\text{s}(\varepsilon-\Delta V^{\text{b}}) \, \eu{-\beta V_0^\text{b}} .
    \label{equ:pi_class3}
\end{equation}
Note that the effect of the convolution manifests itself in a change of the argument of the subsystem lineshape function $I^\text{s}$, which now implicitly depends on the bath coordinate $Q$ via $\Delta V^\text{b}$.

Viewing the expression for the reaction rate with an implicit dependence on the bath coordinates is also the idea that enables the illustration of the MLJ rate by shifted potentials along the bath modes, as shown in Fig.~\ref{fig:pess}. 
In fact, if \eqn{equ:k_fgr} is used for the subsystem lineshape function and the remaining $Q$-integral over the delta-function in \eqn{equ:Ibath} is taken, the standard MLJ rate formula [\eqn{equ:k_bj_full}] is recovered.

Here, we seek to treat the subsystem part with semiclassical instanton theory.
Note that both the MLJ and instanton version of the subsystem lineshape function emerge from \eqn{equ:lss}. The difference is induced by the order in which the sums and integrals in \eqn{equ:lss} are taken. Whereas in MLJ theory the configuration-space integrals are taken before the sums over states are carried out, in instanton theory these steps are taken in reversed order leading to a path-integral instead of a wavefunction formulation of the reaction rate.
Only in the path-integral formulation is it possible to take the steepest-descent integration which leads to semiclassical instanton theory.
The instanton subsystem lineshape function is thus given by %
\footnote{This lineshape function is related to the absorption spectrum calculated in \Ref{GoldenInverted} with an excitation frequency corresponding to $\hbar \omega_{\text{ex}} \equiv \varepsilon - \Delta V^\text{b}$}
\begin{multline}
I^\text{s}_{\text{SCI}}(\varepsilon-\Delta V^{\text{b}}) = \frac{\sqrt{2\pi\hbar}}{Z_0^\text{s}}
\sqrt\frac{C_0 C_1}{C} 
\left( - \frac{\rmd^2 S}{\rmd\tau^2} \right)^{-\frac{1}{2}}\\
\times
\eu{-S(\tau)/\hbar - (\Delta V^{\text{b}} - \varepsilon) \tau/\hbar} ,
\label{equ:Is_sci}
\end{multline}
where again all quantities are evaluated at the stationary point of the exponent $\left[S(\tau)/\hbar + (\Delta V^{\text{b}} - \varepsilon) \tau/\hbar\right]$ in the subsystem coordinates $q'$, $q''$ and imaginary time $\tau$ simultaneously.
Using this approximation in \eqs{equ:pi_class3} and \eqref{equ:pi_class2} defines the instanton formulation of MLJ theory.

Here we show that this approach gives the same result as the reduced instanton theory derived in Sec.~\ref{subsec:rinst_form}. 
By employing \eqn{equ:pi_class3} for the subsystem lineshape function in \eqn{equ:pi_class2}, the effective action in the exponent becomes
\begin{equation}
 \mathcal{S}(Q,\tau) = S(\tau) + (\Delta V^\text{b} - \varepsilon)\tau + \beta\hbar V_0^\text{b} .
 \label{equ:classexp}
\end{equation}
Due to the harmonic nature of the bath, the stationary point in the bath coordinates can be solved for analytically.
This defines the hopping point at which the electron transfer dominantly takes place.
Within the classical limit, it is given by
\begin{equation} \label{Qddag}
    Q^{\ddagger} = \zeta \left(\frac{2\tau}{\beta\hbar} - 1\right) .
\end{equation}
Evaluating \eqn{equ:classexp} at this point therefore leads $\mathcal{S}(Q^\ddagger,\tau) = \mathcal{S}^\text{r}(\tau)$. 
So the exponent becomes identical to that of reduced instanton theory (with a classical bath) and hence the value of $\tau$ at the stationary point is the same too.

The rate expression for this instanton version of MLJ theory is obtained by performing the remaining $Q$-integral by steepest-descent and using the classical partition function $Z_0^\text{b}=(\beta\hbar\Omega)^{-1}$ to give
\begin{align} 
    k_\text{SCI}(\varepsilon)
    &= \frac{\Delta^2}{\hbar^2} \sqrt{\beta\hbar M\Omega^2} \,
     \tilde{I}_\text{SCI} (\varepsilon - \Delta V^{\text{b}}) \left(\der[2]{\mathcal{S}}{Q}\right)^{-\half} ,
    \label{equ:k_imlj}
\end{align}
where all quantities  are evaluated at the stationary point $Q=Q^\ddag$
including the system lineshape function, which implicitly depends on $Q$ through $\Delta V^\text{b}$.

In order to verify the equivalence of this rate expression with \eqn{equ:krinst}, we make use of the rules of consecutive steepest-descent integrations \cite{GoldenGreens,Kleinert}
\begin{equation}
 \der[2]{\mathcal{S}}{Q} = \pder[2]{\mathcal{S}}{Q} - \pders{\mathcal{S}}{Q}{\tau}\left(\pder[2]{\mathcal{S}}{\tau}\right)^{-1}\pders{\mathcal{S}}{\tau}{Q} .
\end{equation} 
Because the spatial subsystem and bath coordinates are independent, the partial derivatives involving $Q$ can be easily evaluated. 
After rearranging, this results in
\begin{equation}
 \pder[2]{S}{\tau}\der[2]{\mathcal{S}}{Q} = %
 \beta\hbar M \Omega^2 \,
 \pder[2]{\mathcal{S}^\text{r}}{\tau} ,
\end{equation}
where $\pder[2]{\mathcal{S}^\text{r}}{\tau} = \pder[2]{S}{\tau} - 2\Lambda/\beta\hbar$.

Using these expressions in \eqn{equ:k_imlj} shows that 
this approach is therefore identical to rSCI [\eqn{equ:krinst}], which is not surprising 
as all we have done is carry out the same steepest-descent integrations but in a different order.

\begin{figure*}
    \centering
    \includegraphics[width=1.0\textwidth]{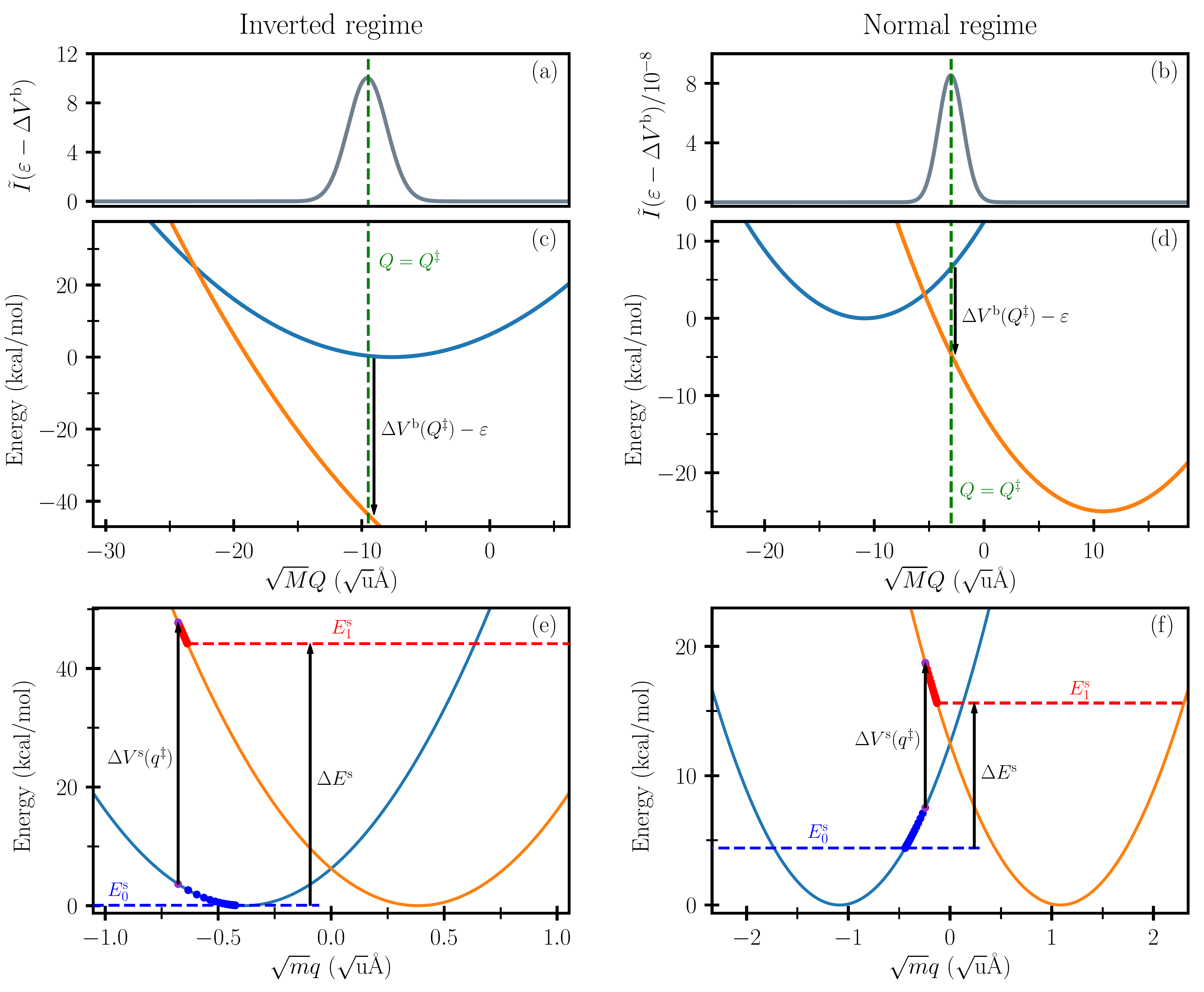}
    \caption{Insights from instanton theory into the reaction mechanism for the inverted and normal-regime model. 
    (a,b) 
    \eqn{equ:pi_class3} as a function of $Q$.
    (c,d) 
    Plot of the potential energy curves (including the driving force, $\varepsilon$) along the bath mode. The location of the hopping point along the classical mode $Q^{\ddagger}$ and the potential energy differences at this point $ \Delta V^\text{b}(Q^\ddagger) - \varepsilon$ are indicated.
    (e,f)
    Plot of the potential energy curves along the subsystem mode together with the optimized ring-polymer instanton corresponding to $Q=Q^{\ddagger}$,
    which was used to compute the subsystem contribution to the rate. The instanton energies in the subsystem [\eqn{equ:Einst}] (dashed lines) and the corresponding energy difference $\Delta E^\text{s}$ are indicated. The energy difference can be measured equivalently as $\Delta V^\text{s}(q^\ddag)$ at the hopping point ($q^\ddag$, purple dot). %
    }
    \label{fig:Integrate}
\end{figure*}

Following this procedure for the displaced harmonic-oscillator models defined in Table~\ref{table:1}, 
we obtain the instantons 
depicted in Fig.~\ref{fig:Integrate}.
Panels (a) and (b) show $\tilde{I}(\varepsilon - \Delta V^{\text{b}})$ computed with the semiclassical instanton approximation as a function of $Q$ for the inverted and normal-regime model.
As expected, the function is centered around $Q^\ddagger$ and is well approximated by a Gaussian.
From instanton theory, we have therefore obtained a reduced picture of the reaction, but this time the focus is along the solvent coordinate and hence can provide a similar interpretation to that from MLJ theory. 

\subsection{Analysis and Mechanistic Insights}
\label{sec:insights}

In addition to the formal connection between SCI and MLJ discussed in Sec.~\ref{subsec:iMLJ},
we will show that, %
as well as the insight into the tunnelling pathway,
it is possible to use instanton theory to extract
very similar information about the reaction as is offered by MLJ theory,
such as the bath activation energy and the
dominant reactant and product vibrational states.
We thus suggest that instanton theory may be used instead of MLJ theory for understanding and interpreting electron-transfer reactions
in complex anharmonic systems.

In Table~\ref{table:2}, we present numerical values of the reaction rates for the two models in the normal and inverted regimes models defined in Table~\ref{table:1}
computed with different methods as well as a number of values obtained from the instanton calculation which we will describe later.
A comparison of the accuracy of the approaches has already been carried out in Sec.~\ref{subsubsec:morse} and thus here we simply note a couple of points which are special to this case.
The fact that for the inverted-regime model the rSCI rate is even slightly closer to the quantum rate than the MLJ rate can be attributed to a fortuitous error cancellation, as MLJ theory is, in principle, the more accurate method in this case.
Because both models consist of displaced symmetric harmonic oscillators, the second-order cumulant expansion is exact in these cases and therefore not shown.
In contrast, the rate obtained with the SFC method, while showing decent agreement with the exact rate for the inverted-regime model, exhibits an error of almost one order of magnitude for the normal-regime model. This is in agreement with the findings in \Refs{Siders1981quantum,Siders1981inverted}.

Just as in the reduced instanton formalism derived in Sec.~\ref{subsec:inst_theory}, the instantons computed in the subsystem coordinate space, shown in the bottom panels of Fig.~\ref{fig:Integrate}, consist of paths $q_n(u)$ whose energies are not equal but differ by the amount $\Delta E^\text{s}$.
We will show that this energy jump is a good approximation to the difference in energies between the dominant reactant and product vibrational states in the MLJ sum, i.e.\ $E_1^\nu-E_0^\mu$.
As explained in \Ref{GoldenInverted},
in the normal regime, the trajectories travel in opposite directions away from the hopping point $q^\ddag$,
but in the same direction when in the inverted regime. This occurs because $\tau<0$ in the inverted regime such that the product trajectory travels in negative imaginary time and thus in the opposite direction from its momentum.

The energy jump in the bath is indicated in Figs.~\ref{fig:Integrate}(c) and (d) and can be defined from   the potential energy difference at the hopping point [\eqn{Qddag}]
\begin{equation}
    \Delta E^\text{b} = \Delta V^\text{b}(Q^\ddag) = \Lambda^\text{b} \left( 1 - 2 \frac{\tau}{\beta\hbar} \right) ,
\end{equation}
which is seen to be equal to $\pder{\Phi_\text{cl}}{\tau}$ and just justifies identifying this term as
the energy jump in the bath in Sec.~\ref{subsec:rinst_form}.
At the stationary point we have $\Delta E^\text{s} + \Delta E^\text{b} - \varepsilon=0$, which confirms that the total energy is conserved.

As well as predicting the energy jump, we can also predict the reactant and product vibrational states which dominate the MLJ sum.
In instanton theory the energies of the two trajectories $q_n(u)$ making up the reduced instanton, defined by \eqn{equ:Einst}, indicate the energies with the largest contributions to the thermal rate. %
In this harmonic system we can relate the energies directly to the vibrational quantum numbers as the energy levels are known.
This would of course not be possible in a complex system, although knowledge of the energy in the subsystem before and after the reaction, which provides similar insight, would still be available.

\begin{table}
\caption{Computed quantities for the harmonic models defined in Table~\ref{table:1}. The reduced instanton rates with classical bath and the corresponding values $\tau$ at the stationary point of the reduced action $\mathcal{S}^\text{r}$ were obtained from ring-polymer instanton optimizations with $256$ beads equally distributed between both electronic states. The contributions to the total effective action from subsystem $S_n$ and bath $\Phi_\text{cl}$ are also given.
The exact rate is obtained from integration of the flux correlation function of the full system.\cite{GoldenGreens} 
As described in Appendix~\ref{app3}, the SFC approximation is used for both subsystem and bath in order to compute the corresponding rates.
For both models, the MLJ rate includes contributions from excited reactant states.
All rates, including the Marcus rate for the full system $k_\text{MT}(\varepsilon)$, are given relative to the Marcus rate for the bath of the respective model only $k_\text{MT}^\text{b}(\varepsilon)$.
}
\label{table:2}
\begin{ruledtabular}
\begin{tabular}{lcc} 
 & Inverted-regime & Normal-regime\\
 & model & model\\
 \hline
 $E_0^{\text{s}}/\hbar\omega$  & $0.02$ & $3.08$\\ 
 $E_1^{\text{s}} /\hbar\omega$  & $15.46$ & $10.91$\\
 $\tau/\beta\hbar$ & $-0.12$ & $0.36$ \\
 $S_0/\hbar$ & $2.538$ & $6.787$ \\
 $S_1/\hbar$ & $-9.171$ & $10.700$ \\
 $\Phi_\text{cl}/\hbar$ & $-5.501$ & $19.367$ \\
 $V^\text{b}_{0}(Q^\ddag)$  (\si{\Calorie\per\mol}) & $0.34$ & $6.55$ \\ 
\hline
 $k_\text{ex}(\varepsilon)/k_\text{MT}^\text{b}(\varepsilon)$ & $5.159\cdot 10^{16}$ & $5.484\cdot 10^{-8}$ \\
  $k_\text{MLJ}(\varepsilon)/k_\text{MT}^\text{b}(\varepsilon)$ & $5.144\cdot 10^{16}$ & $5.366\cdot 10^{-8}$ \\
 $k_\text{rSCI}(\varepsilon)/k_\text{MT}^\text{b}(\varepsilon)$ & $5.152\cdot 10^{16}$ &  $5.360\cdot 10^{-8}$ \\ 
   $k_\text{SFC}(\varepsilon)/k_\text{MT}^\text{b}(\varepsilon)$ & $4.221\cdot 10^{16}$ & \!\!$49.856\cdot 10^{-8}$ \\
 $k_\text{MT}(\varepsilon)/k_\text{MT}^\text{b}(\varepsilon)$ & $0.615\cdot 10^{16}$ & $0.762\cdot 10^{-8}$ \\ 
\end{tabular}
\end{ruledtabular}
\end{table}

As one can read from the table, for the inverted-regime model, %
the instanton energies correspond to a transition from the reactant ground vibrational state $\mu\approx0$ to the product state $\nu\approx15$.
The dominant vibrational channel in the normal-regime model is predicted to involve an excited reactant vibrational state $\mu\approx3$ and the product state $\nu\approx11$.
A comparison of these values with Figs.~\ref{fig:pess}(c) and \ref{fig:pess}(d) reveals that, for both models, the instanton energy picks out the same dominant vibrational channel as MLJ theory.
Note that instanton theory does not actually quantize the reactant and product wells as it relies solely on imaginary-time trajectories which exist only in the classically forbidden regions.
It does not therefore give integer values for the dominant states. %
This is however not a serious concern as there is no particular relevance of the individual state with the largest contribution because typically MLJ theory predicts that a cluster of states are involved and thus any prediction within the cluster is practically as good.
\footnote{There is also no problem that the predicted energy is lower than the zero-point energy.
In fact, semiclassical trajectories predict the exact partition function of the harmonic oscillator at any temperature despite having an energy of 0 \cite{InstReview}}

Furthermore, for a subsystem in conjunction with a classical harmonic bath,
the activation energy of the bath modes can be easily recovered using \eqn{Qddag} to give
\begin{align}
    \label{equ:bathact}
    V_0^\text{b}(Q^\ddag) &= \Lambda^{\text{b}} \left(\frac{\tau}{\beta\hbar}\right)^2 ,
\end{align}
which should be evaluated at the stationary value of $\tau$.
The values for the bath activation energy obtained from rSCI theory are also given in Table~\ref{table:2} and are in good agreement (i.e.\ with an error less than the thermal energy) with the results obtained from MLJ in theory given in Sec.~\ref{subsubsec:models}.

In the inverted-regime model, as can be seen in Fig.~\ref{fig:Integrate}(c), the bath activation energy is thus substantially lower than
it would be if there were no subsystem, for which it would correspond to the point where the potentials cross.
The presence of the subsystem therefore leads to a significant speed-up of the reaction, which explains why the rate for the full system in Table~\ref{table:2} is many orders of magnitude larger than the corresponding reaction taking place in the bath only. 
However, the rate is not only dependent on the bath activation energy but also depends on the action of the subsystem instanton, as can be seen from \eqn{equ:classexp}.
As previously discussed, the stationary value of the bath configuration, $Q^\ddag$,  is associated with an energy jump $\Delta V^\text{b}(Q^\ddag)-\varepsilon$, that  must be compensated by $\Delta E^s=\Delta V^\text{s}(q^\ddag)$ with an equal magnitude but opposite sign in the subsystem in order to satisfy energy conservation.
Panels (c) and (e) of Fig.~\ref{fig:Integrate} illustrate that minimizing the bath activation energy causes the tunnelling pathway in the subsystem to lengthen which increases the system action. The elongation of the path can be understood from Fig.~\ref{fig:Integrate}(e) which shows that, in order to reach a point where the potential-energy difference between the subsystem potentials exactly compensates the energy jump in the bath, the system has to travel ``uphill'' to the left. Hence, in general a compromise has to be made between minimizing the bath activation energy and the subsystem action.
Only in one particular case %
is the magnitude of the energy jump at the reactant minimum in the bath (including the driving force) %
identical to the potential-energy difference at the reactant minimum in the subsystem
such that an activationless reaction becomes possible.  %
In this special case, the energy jump in the bath is $\Lambda^\text{b}-\varepsilon$ and in the system is $\Delta V^\text{s}(q_\text{min}^{(0)})$,
which %
for our example where $V_0^\text{s}(q_\text{min}^{(0)}) = V_1^\text{s}(q_\text{min}^{(1)})$,
leads to the requirement
$\varepsilon=\Lambda$. %
The rates in the inverted regime are much faster than in the fully classical treatment, because the reaction within the subsystem can proceed via quantum tunnelling, as depicted in Figs.~\ref{fig:Integrate}(e) and (f), instead of relying on thermal activation. 
Altogether, this implies that, although the turnover curve in the inverted regime is not as steep as it would be according to the classical theory, it does not become independent of $\varepsilon$.

On the other hand, in the normal regime, the bath activation energy in Fig.~\ref{fig:Integrate}(d) is seen to be higher than
it would be without the presence of the subsystem.
This causes the rate for the full system to decrease relative to the electron-transfer reaction in the bath only.
The tunnelling effect increases the rate relative to a classical calculation, although typically not as dramatically as in the inverted regime.  This can also be understood from an analysis of the instanton tunnelling trajectories as was explained in \Ref{GoldenInverted}.

Thus, we were able to show how practically all insights from MLJ theory including, first and foremost, the dominantly contributing reactant and product energies can equally be obtained from reduced instanton theory.  Semiclassical instanton theory further allows one to attain this understanding of the reaction even in complex, anharmonic systems. This information is complemented by the localization of the optimal tunnelling pathway in the subsystem, which can be interpreted as the reaction mechanism in configuration space.

\section{Conclusions}

We have developed an instanton formulation of MLJ theory, which focuses on the subsystem while including a classical or quantum harmonic bath implicitly.
This provides a practical method to complement the
simulation of electron-transfer reactions of multidimensional anharmonic subsystems by the effect of a solvent environment.
Thus, the method is ideally suited to study problems that necessitate multiscale modelling. 
Electron-transfer rates have been calculated and compared to results from several other methods for an asymmetric anharmonic model %
and the results demonstrate that reduced semiclassical instanton theory is in excellent agreement with 
either the exact rate 
or the MLJ rate depending on whether the bath is assumed to be classical or not.
Thus we argue that semiclassical instanton theory can be reliably employed in situations which have previously been simulated by MLJ theory.

In addition to MLJ theory, we have also compared our approach to 
the second-order cumulant expansion,
a popular method commonly used to describe electron-transfer and optical transition rates, and to the semiclassical Franck--Condon sum.
The results obtained with both these approximations exhibit severe errors, especially in the normal regime and in fact, unlike instanton theory, 
neither the cumulant expansion nor the SFC approximation satisfy the detailed balance relation [\eqn{equ:detbal}].
This underlines the fact that, although both the SCI and SFC methods have been termed ``semiclassical'', %
the approximations are quite unrelated.

We also compared and contrasted the insight that MLJ and instanton theories can offer into the mechanism of electron-transfer reactions.
The traditional MLJ picture is shown along the bath coordinates, in which the subsystem has an effect by shifting the reactant and product potential by their respective internal energy levels. %
Although undoubtedly simple and intuitive in one dimension,  this picture quickly becomes convoluted when a multidimensional anharmonic subsystem has to be considered.
There is also little insight given into the tunnelling dynamics of the subsystem itself.

Instanton theory, on the other hand, automatically locates a unique reaction coordinate which describes the optimal tunnelling pathway of the subsystem modes.
In analogy to the dynamics of open quantum systems,
the addition of a bath changes the instanton pathway in the subsystem such that, due to energy exchange between subsystem and bath, the reduced instanton exhibits an additional jump in energy at the hopping point.
Although the energy in the subsystem is therefore not conserved by the electron-transfer reaction,
the excess energy is absorbed by the bath such that the total energy is conserved as it of course should be.
This picture of tunnelling under the barrier along a reaction coordinate reflects the typical situation of practical simulations,
where the focus is on the subsystem under the influence of a surrounding solvent bath.

Nonetheless, we have also discussed how instanton theory is connected to MLJ theory by deriving them both from a common expression.
This shows that in principle similar insights can be extracted from either method.
In particular, we show that instanton theory can successfully predict the same dominant initial and final vibrational state of the system before and after the electron-transfer event as MLJ theory.

Instanton theory overcomes the main disadvantage of MLJ theory, which is that it requires knowledge of the energy levels and wavefunctions of the subsystem.
Because of this, applications of MLJ theory are often limited to a harmonic-oscillator approximation, which introduces an uncontrolled error when simulating an anharmonic system.
Hence, although rSCI (with a classical bath) is technically an approximation to MLJ theory, %
in many anharmonic cases it will lead to more accurate results due to its ability to account for anharmonicity along the tunnelling pathway. %
In conjunction with a ring-polymer discretization,
instanton theory can be applied directly to multidimensional anharmonic problems.
The application of this theory to electron-transfer reactions, spin transitions and energy-transfer processes of molecular systems
in combination with high-level \textit{ab-initio} electronic structure methods, as has been used in previous instanton studies,\cite{porphycene,GPR,dimersurf} 
will be integral part of future work.

\section*{Acknowledgements}
This work was financially supported by the Swiss National Science Foundation through SNSF Project 175696.

\appendix

\section{Quantum and classical rate formulas for an anharmonic subsystem mode in conjunction with a harmonic bath}
\label{app1}

In order to put the instanton and MLJ results shown in Fig.~\ref{fig:rates300} into context, we also present the exact quantum rates for this system and their classical limits. This enables us not only to directly check the quality of the results obtained with the approximate methods, but by comparison with the classical rates also allows an estimation of the relevance of nuclear quantum effects.
In this section, we harness the formal framework laid out in Sec.~\ref{sec:general} to derive the required rate formulas making use of the fact that we can analytically integrate out the coordinate-dependence of the harmonic bath.

If the wavefunctions of the subsystem are known, as is the case for the two crossing Morse oscillators used in Sec.~\ref{subsubsec:morse}, the trace over the subsystem degrees of freedom in \eqn{equ:lss_gen} can be evaluated exactly in the wavefunction representation. %
Thus, the exact quantum-mechanical rate can be computed by the formula
\begin{multline}
    k(\varepsilon) \, Z_0^{\text{s}} = \frac{\Delta^2}{\hbar^2}
    \int_{-\infty}^{\infty} \rmd t \SumInt_\mu \eu{-\beta E_0^{\mu}}
    \SumInt_\nu |\theta_{\mu\nu}|^2\\
    \times
    \eu{-(\tau + \iu t)(E_1^\nu - E_0^\mu - \varepsilon)/\hbar - \Phi(\tau + \iu t)/\hbar } 
     ,
    \label{equ:k_exact}
\end{multline}
where sums are taken over the bound states of the Morse potential and integrals are carried out over the energies of the energy-normalized continuum states.
Expressions for the wavefunctions can be found in \Refs{Muendel1984morse,Bunkin1973Morse}.
Numerical integration was used to obtain the Franck--Condon overlaps and to perform the integral over time.
In order to make the latter converge easily, 
the imaginary-time variable $\tau$ was chosen appropriately (i.e.\ using the value obtained from the instanton optimization).

The classical limit of this rate can be obtained in a similar way except that the trace in \eqn{equ:lss_gen} is evaluated by a classical phase-space integral\cite{Schmidt1973} and the classical limit of the effective bath action is used [\eqn{equ:SBclassS}].
For a one-dimensional subsystem, this gives the  %
classical golden-rule transition-state theory rate
\begin{multline}
    k_{\text{TST}}(\varepsilon) \, Z_0^{\text{s}} = \frac{\Delta^2}{\hbar^3} \sqrt{\frac{m}{2\Lambda^{\text{b}}}}
    \int \eu{- \beta V_0^{\text{s}}(q)}\\ 
    \times  \eu{-\beta ( \Lambda^{\text{b}} - \varepsilon + \Delta V^{\text{s}}(q))^2/4\Lambda^{\text{b}}} \rmd q ,
    \label{equ:TST}
\end{multline}
where %
the reactant partition function is computed by a classical phase-space integral.
The remaining integral in the subsystem mode can either be taken numerically, as was done to generate the results in Fig.~\ref{fig:rates300}, or by steepest descent, which would be an excellent approximation in this case.

The quantum-mechanical rate [\eqn{equ:k_exact}], as well as the MLJ rate [\eqn{equ:k_bj_full}] correctly reduce to the TST expression in \eqn{equ:TST} in the high-temperature or low-frequency limit, while instanton theory [\eqn{equ:krinst}] reduces to the steepest-descent version of it.

In the special case that the subsystem consists of displaced harmonic oscillators, \eqn{equ:TST} reduces to the Marcus theory expression [\eqn{equ:marcus}] except that in this case the reorganization energy should be the sum of the subsystem and bath reorganization energies.

\section{Cumulant expansion}
\label{app2}

Another approximate way of computing correlation functions and therefore also to calculate electron-transfer reaction rates is the so called ``cumulant expansion'',\cite{Kubo1962,Hashitsume1998,Loring1987} which in our formulation will be applied to the lineshape function of the subsystem \eqn{equ:lss_gen}.

In its conventional formulation $\tau$ is set to zero and we rewrite \eqn{equ:lss_gen} as
\begin{equation}
    I^\text{s}(v) = 
    \! \int_{-\infty}^\infty
    \eu{\iu v t/\hbar} \,
    R(t)
    \, \rmd t ,
    \label{equ:cumulant1}
\end{equation}
where the correlation function is
\begin{equation}
     R(t) = \left(Z_0^\text{s}\right)^{-1} 
    \!
    \Tr_{\text{s}}\big[\eu{-(\beta\hbar - \iu t )\op{H}_0^\text{s}/\hbar}
    \, \eu{-\iu t \op{H}_1^\text{s}/\hbar}\big] .
    \label{equ:cumulant2}
\end{equation}
The time-dependent terms inside the trace can equally be rewritten as a time-ordered exponential according to
$\eu{+\iu t\op{H}_0^\text{s}/\hbar}\,\eu{-\iu t\op{H}_1^\text{s}/\hbar} = \op{\mathcal{T}} \eu{-\iu\int_0^t \Delta\op{V}^\text{s}_\text{I}(t') \rmd t'/\hbar}$
where $\op{\mathcal{T}}$ is the time-ordering operator 
and we make use of the interaction picture to give
$\Delta \op{V}^\text{s}_{\text{I}} (t) = \eu{+\iu \op{H}_0^\text{s} t/\hbar} \, \Delta \op{V}^\text{s} \, \eu{-\iu \op{H}_0^\text{s} t/\hbar}$,
where
$\Delta \op{V}^\text{s} = \op{H}^\text{s}_1 - \op{H}^\text{s}_0 = V^\text{s}_1(\op{q}) - V^\text{s}_0(\op{q})$.
\cite{KuehnBook}
This exact expression can then be expanded in a time-ordered power series with respect to $\Delta \op{V}^\text{s}_\text{I}$.

Motivated by the analytic solution for the correlation function of a system of displaced harmonic oscillators [\eqn{equ:lsb}], one makes the ansatz $R(t) = \exp\left[-\Gamma(t)\right]$ %
where the exponent is defined as a sum of cumulants
\begin{equation}
    \Gamma(t) = \sum_{j=1}^\infty \Gamma_j(t) ,
    \label{equ:gamma}
\end{equation}
where $\Gamma_j(t)$ is of $j$th order in $\Delta \op{V}^\text{s}_\text{I}$.
Comparing the two expansions, the first two terms in \eqn{equ:gamma} are given by
\cite{KuehnBook}
\begin{subequations} \label{cumulants}
\begin{align}
    \Gamma_1(t) &=  \frac{\iu}{\hbar} \left(Z_0^\text{s}\right)^{-1}  \int_0^{t} \rmd t_1 
    \Tr_{\text{s}}\big[ \eu{- \beta \op{H}_0^\text{s}} \Delta \op{V}^\text{s}_{\text{I}} (t_1)  \big] ,\\ 
    \nonumber
    \Gamma_2(t) &= \thalf \Gamma_1^2(t)
    + \frac{1}{\hbar^2} \left(Z_0^\text{s}\right)^{-1}\\
    &\times \int_0^{t} \rmd t_1 \int_0^{t_1} \rmd t_2 
    \Tr_{\text{s}}\big[ \eu{- \beta \op{H}_0^\text{s}} \Delta \op{V}^\text{s}_{\text{I}} (t_1) \Delta \op{V}^\text{s}_{\text{I}}(t_2)  \big] ,
\end{align}
\end{subequations}
whereas higher cumulants are neglected in the expansion.
The expressions in \eqs{cumulants} can be evaluated by expanding the traces in the energy-eigenstate basis of $\op{H}_0^\text{s}$. Performing the time-integrals analytically results in the equations
\begin{subequations}
\label{equ:gammas}
\begin{align}
    \Gamma_1(t) &= \frac{\iu t}{\hbar} \left(Z_0^\text{s}\right)^{-1}   
    \sum_{\mu} \eu{- \beta E^\mu_0}  \Delta V^\text{s}_{\mu\mu} ,\\ 
    \nonumber
    \Gamma_2(t) &= \thalf \Gamma_1^2(t)
    + \left(Z_0^\text{s}\right)^{-1} \sum_\mu \sum_{\mu'}
    \eu{- \beta E^\mu_0} |\Delta V^\text{s}_{\mu\mu'} |^2 \\
    \label{equ:Gamma2}
    &\times  \frac{1 
    + \iu (E_0^\mu - E_0^{\mu'})t/\hbar
    - \eu{\iu (E_0^\mu - E_0^{\mu'}) t/\hbar} }{(E_0^\mu - E_0^{\mu'})^2}
    ,
\end{align}
\end{subequations}
where $\Delta V^\text{s}_{\mu\mu'} = \int_{-\infty}^{\infty} \psi_0^{\mu}(q)^* \Delta V^\text{s}(q)    \psi_0^{\mu'}(q) \, \rmd q$ and these integrals over the one-dimensional subsystem coordinate
are evaluated numerically. The terms in the sum of \eqn{equ:Gamma2} with $\mu=\mu'$ can be evaluated by L'Hôpital's rule. 
For the case of displaced harmonic oscillators, this expansion of the correlation function up to second order gives the exact result, as all higher order terms vanish.\cite{KuehnBook}
In the general, anharmonic case, however, the quality of the approximation is unclear.
One could of course extend the method to higher orders, but the series is unlikely to converge quickly to correct result.

The computational advantage of the cumulant expansion over the golden-rule formula is that only the eigenstates of the reactant electronic state need be known.
It is thus perhaps most useful when computing absorption spectra from a ground electronic state to an excited state, for which the ground state is well approximated by a harmonic oscillator, but not the excited state.
However, a significant knowledge of the product potential-energy surface is still required in the region 
where the wavefunction overlaps in \eqs{equ:gammas} are sizeable, which can be expensive to compute. 

This result for the subsystem's lineshape function can be easily combined with the lineshape function of the harmonic bath from \eqn{equ:lsb} by performing the convolution integral in $v$ and integrating over the resulting delta function.
The rate expression based on the second-order cumulant expansion for the subsystem part is therefore given by 
\begin{multline}
    k_{\text{CE}}(\varepsilon) = \frac{\Delta^2}{\hbar^2}
    \int_{-\infty}^{\infty}  
    \eu{\iu \varepsilon t /\hbar} \,
    \eu{- \Gamma_1(t) - \Gamma_2(t) - \Phi(\iu t)/\hbar } 
    \, \rmd t
     ,
    \label{equ:k_ce}
\end{multline}
where either a quantum or classical bath can be employed by using the respective expressions for the actions in \eqs{equ:Sspinboson} and \eqref{equ:SBclassS} and the final time-integral is carried out numerically.

Note that this cumulant expansion leads to a completely different approximation from that of Wolynes theory \cite{Wolynes1987nonadiabatic} even though the latter can also be thought of as a type of cumulant expansion.
In contrast to the approach described here, Wolynes theory carries out the time integral by the method of steepest-descent and computes the short-time limit of the correlation function by path-integral sampling.
For the systems studied in this work, Wolynes theory would give similar results to those of instanton theory (identical in the case of a harmonic system), although for certain more complex systems it has been shown to break down.\cite{nonoscillatory,GRQTST,GRQTST2,Fe2Fe3}
Unlike the cumulant expansion and instanton theory,\cite{GoldenInverted} it is also not directly applicable to the inverted regime, although an extrapolation method which extends it in this way has been suggested. \cite{Lawrence2018Wolynes}

\section{Semiclassical Franck--Condon Sum}
\label{app3}

The ``semiclassical Franck--Condon sum'' is an alternative way of approximating the electron-transfer rate
and can be obtained from \eqn{equ:k_et} by neglecting the commutator between $\op{H}_0$ and $\op{H}_1$ in both subsystem and bath, setting $\tau$ to zero and evaluating the trace in the reactant's eigenfunction basis.\cite{Schmidt1973,Siders1981quantum,Marcus1984semiclassical} 
For the lineshape function of the one-dimensional subsystem, this results in
\begin{equation}
    I_\text{SFC}^\text{s}(v) = \frac{2\pi\hbar}{Z_0^\text{s}} \sum_\mu \eu{-\beta E_0^\mu} \int \psi_0^\mu (q)^* \,
    \delta(\Delta V^\text{s}(q) - v)
    \,  \psi_0^\mu (q) \, \rmd q
    , 
\end{equation}
where by virtue of neglecting the commutators, we were able to make the classical approximation $\int\eu{\iu\op{H}^\text{s}_0t/\hbar}\,\eu{-\iu (\op{H}^\text{s}_1 - v) t/\hbar} \, \rmd t \approx 2\pi\hbar \, \delta(\op{H}^\text{s}_1-\op{H}^\text{s}_0 - v)$.  Because the kinetic part vanishes in the difference of the Hamiltonians the final expression can be written in terms of $\Delta V^\text{s} (q) \equiv V_1^\text{s}(q) - V_0^\text{s}(q)$.

The same strategy is used to deal with the bath.
However, as described in the appendix of \Ref{Siders1981quantum}, because the bath is harmonic,
the sums and integrals can be performed analytically to give
\begin{equation}
    I_\text{SFC}^\text{b}(\varepsilon - v) = \sqrt{\frac{\pi\beta\hbar^2}{\chi\Lambda^\text{b}}}  
    \, \eu{- \beta(\Lambda^\text{b} - \varepsilon + v)^2 / 4 \chi \Lambda^\text{b}} ,
    \label{equ:sfc_bath}
\end{equation}
which has the same form as that of Marcus theory except for the
correction factor, $\chi = \sum_{j=1}^{D} \Lambda^\text{b}_j \gamma_j \coth{\gamma_j} / \Lambda^\text{b}$,
which is defined in terms of the reorganization energy associated with a single bath mode $\Lambda^\text{b}_j = 2M\Omega_j^2\zeta_j^2$ and $\gamma_j = \beta \hbar \Omega_j / 2$.

Following the formalism laid out in Sec.~\ref{sec:general} and performing the convolution integral in $v$ first, this leads the rate equation
\begin{multline}
    k_\text{SFC}(\varepsilon) \, Z_0^\text{s} = \frac{\Delta^2}{\hbar} 
    \sqrt{\frac{\pi\beta}{\chi\Lambda^\text{b}}}
    \sum_\mu \eu{-\beta E_0^\mu}\\
    \times
    \int \psi_0^\mu (q)^* \, \eu{- \beta (\Lambda^\text{b} - \varepsilon + \Delta V^\text{s} (q))^2 / 
    4 \chi\Lambda^\text{b}} \,
    \psi_0^\mu (q) \, 
    \rmd q ,
    \label{equ:sfc}
\end{multline}
where the integrals over the anharmonic subsystem mode have to be carried out numerically.

In the special case in which all modes are displaced harmonic oscillators, all degrees of freedom can be assigned to the bath. Then, the rate formula is directly given by $k_\text{SFC}(\varepsilon) = \frac{\Delta^2}{\hbar^2} I_\text{SFC}^\text{b}(\varepsilon)$.   
It is easy to see that this is in error because it predicts results symmetric around $\varepsilon=\Lambda$, whereas the true result is known to be significantly skewed unless in the classical limit.\cite{Siders1981inverted,Ulstrup1975,Bixon1991ET}
One way to understand the causes of this error has been explained in terms of WKB theory in \Ref{Marcus1982a}.

\bibliography{references,other}
\end{document}